\documentclass[aps,prb,twocolumn,subscriptaddress,floatfix]{revtex4-1}
\usepackage[parfill]{parskip}   
\usepackage{graphicx}
\usepackage{amssymb}
\usepackage{amsmath}
\usepackage{color}

\begin{document}

\title{Relaxation of electrons in quantum-confined states in Pb/Si(111) thin films from master equation with first-principles-derived rates}
\author{Peter Kratzer and Maedeh Zahedifar}
\affiliation{
Fakult\"at f\"ur Physik, Universit\"at Duisburg-Essen, Campus Duisburg,
Lotharstr.~1, 47057 Duisburg, Germany
}

\begin{abstract}
Atomically thin films of Pb on Si(111) provide an experimentally tunable system comprising a highly structured electronic density of states. 
The lifetime of excited electrons in these states is limited by both electron-electron (e-e) and electron-phonon (e-ph)  scattering.
We employ the description by a master equation for the electronic occupation numbers to analyze the relative importance of both scattering mechanisms. 
The electronic and phononic band structures, as well as the 
matrix elements for electron-phonon coupling within deformation potential theory were obtained from density functional calculations, thus taking into account quantum confinement effects. For the relaxation dynamics, 
the contribution of impact ionization processes to the lifetime is estimated from the imaginary part of the electronic self-energy calculated in the $GW$  approximation. By numerically solving rate equations for the occupations of the Pb-derived electronic states coupled to a phononic heat bath, we are able to follow the distribution of the electronic excitation energy to the various modes of Pb lattice vibrations. 
While e-e scattering is the dominant relaxation mechanism, we demonstrate that the e-ph scattering is highly phonon-mode-specific, with a large contribution from surface phonons. At electron energies of about 0.3~eV above the Fermi surface, a 'phonon bottleneck' characteristic of relaxation in nanostructures with well-separated electronic states is observed. 
The time scales extracted from the simulations are compared to data from pump-probe experiments using  time-resolved two-photon photoemission.
\end{abstract}
\maketitle
\section{Introduction}
The thermalization of hot carriers in metals after optical excitation is accomplished both by the Coulomb scattering among the carriers (electron-electron (e-e) interactions) and by the scattering of electrons and holes by lattice vibrations (electron-phonon (e-ph) interaction). In a well-established picture~\cite{FaSt92,GrSp95}, the relaxation can be understood as a two-step process: at early times ($t < 0.3$~ps), e-e scattering dominates and brings the electrons to a thermal (or possibly non-thermal) distribution. At later times ($> 0.3$~ps) the e-ph interaction establishes equilibrium between the electronic distribution and the lattice temperature. In this second stage, the high density of excited carriers close to the Fermi energy (within an energy interval corresponding to few phonon quanta) is thought to be responsible for most of  the energy flow between the electronic and the phononic system. If so, the e-ph coupling inferred from thermalization experiments should relate directly to the microscopic e-ph coupling constant that governs electric resistivity or the superconducting transition temperature \cite{Allan87}. 
In this prevailing view, the role of e-ph interactions already in the {\em early} stages of relaxation is usually ignored. 
However, this simple picture is questioned by studies, both experimental and theoretical \cite{LiLo04,BaKa14}, suggesting overlapping timescales of e-e and e-ph-driven thermalization. 
Moreover, there is little knowledge how the electrons {\em far above} the Fermi level (several tenth of eV) interact with the phonons. For instance, long-lived  population of such states, e.g. at the Pb-covered Si(111) surface, has been observed in photoemission experiments \cite{SaMi15}.  The situation at high energies is in contrast to the e-ph interaction in close vicinity to the Fermi surface, which is crucial for a variety of physical phenomena such as electrical resistivity or superconductivity induced by electron-phonon coupling in thin films \cite{ZhCh10}, and is quite well explored utilizing the concept of the Eliashberg function (for an overview, see Ref.~\onlinecite{Hofmann2009}). 
In conclusion, there is a need for more studies of the e-ph interaction at energies further away from the Fermi energy.

In this paper, we attempt to obtain a better understanding of the relative importance of e-e  and e-ph interaction in highly excited states of a metal and their respective contributions to the early stage of relaxation. To introduce our approach, we have chosen thin multilayer Pb films on Si(111). The fact that this materials system shows a highly structured electronic density of states due to confinement effects \cite{ZhJi05,UpWe07} has been a great advantage for analyzing the energy-dependent lifetime of the excited electrons using time-resolved pump-probe spectroscopy.\cite{Kirchmann2010} The experimental results were rationalized in Ref.~\onlinecite{Kirchmann2010} in terms of e-e interaction only, and it was concluded that the electronic lifetime closely follows the behavior expected from Landau's theory of Fermi liquids.\cite{Echenique2004,Chulkov2006} 
Yet, a contribution of e-ph scattering to the lifetime cannot be excluded completely based on the achieved level of agreement between experiment and theory.
Therefore, we aim at a detailled analysis of the role of e-ph scattering for the features observed in photoemission.

Since ample experimental and computational data are available for the Pb/Si(111) films, we consider this system a good test case for quantitative studies of electronic relaxation dynamics.
In a previous paper~\cite{ZaKr17} by us, we have worked out a realistic atomistic description for multilayer Pb films on Si(111) and have carried out first-principles calculations of the electronic and phononic band structure and of e-ph coupling in electronic states far away from the Fermi level. 
While the e-ph interaction in bulk solids has become accessible to first-principles calculations 
by using density functional perturbation theory together with Wannier interpolation methods to enhance the number of reciprocal-space sampling points \cite{Gius17,BeVi14,JhZh17,MaCa17}, thin films on a substrate are still difficult to treat on a microscopic level 
because the adequate supercell typically contains tens to hundreds of atoms and computational costs are high. 
For the Pb films on Si(111), for instance, 
the complex phase diagram \cite{ChWa03,YaYe04} results in various reconstructions requiring large supercells for their description \cite{ SoCh15}. 
In the present work, we constructed a $\sqrt 3 \times \sqrt 3$ unit cell of Si(111) matched with a $(2 \times 2)$ unit cell of Pb(111) to describe the atomic structure consisting of 40 Pb and 30 Si atoms.\cite{ZaKr17} While the two-dimensional Brillouin zone of reconstructed surface plus interface is smaller than the Brillouin zone of a bulk material, the supercell contains a large number of bands, both in the electronic and phononic spectra. Therefore, a thoughtful selection of bands will be required to arrive at a tractable model for e-ph coupling. 
The approach via density functional perturbation theory and the calculation of the Eliashberg function would be too cumbersome for large supercells.

In this paper, building upon the knowledge of our previous work \cite{ZaKr17}, we elaborate on the consequences of these microscopic data for the e-ph scattering rate using a kinetic master equation. The detailed modeling of e-ph scattering is combined with a description of the e-e interaction at the level of Fermi liquid theory. This combination allows us to simulate the temporal evolution of electronic populations on the relevant scales and to make contact with experimental observations.

\section{Theory}
The general problem of an excited electronic system coupled to lattice degrees of freedom can be 
approached from various perspectives. If one is satisfied with a classical description of the atomic positions and velocities and their dynamics
can be described in the trajectory approximation, carrying out non-adiabatic molecular dynamics simulations (e.g. with the methodology described in \cite{AkPr13,AkPr14}) is the method of choice. As an advantage, this approach can handle large deviations of the atomic positions from their ground state, and the forces acting on the atoms are calculated directly within the first-principles electronic structure framework. Thus, it is suitable for systems with 
very strong and non-linear electron-phonon coupling, as encountered  e.g. in two-dimensional materials \cite{Zheng17}. 
In this work, we emphasize the quantum nature of the phonons, while the weak coupling of the electrons to phonons and to external fields can be treated in first-order perturbation theory. 
Casting the problem into the form of a model Hamiltonian, it reads $H =  H_0 + H_{\rm int}$ with $H_0$ being the ground-state Hamiltonian with phonons described in the harmonic approximation, 
\begin{equation}
H_0  =  \sum_{n\mathbf{k}} \varepsilon_{n\mathbf{k}} \, c^{\dagger}_{n\mathbf{k}}
c_{n\mathbf{k}} + \sum_{I \mathbf{Q}} \hbar\Omega_{I\mathbf{Q}} \, 
b^{\dagger}_{I\mathbf{Q}} b_{I\mathbf{Q}} \, .
\end{equation}
The creation and annihilation operators $c_{n\mathbf{k}}, \, c^{\dagger}_{n\mathbf{k}}$ and $b_{I\mathbf{Q}} , \, b^{\dagger}_{I\mathbf{Q}}$ obey the usual anticommutator relations for fermions and commutator relations for bosons, respectively. 
The first, integer index $n$ specifies the band, while the second index $\mathbf{k}$ describes the crystal momentum in the form of a two-dimensional vector within the Brillouin zone of a thin slab. 
Capital letters are used to index phonon modes, whereas small letters refer to electronic bands. 
In contrast to the molecular dynamics approach mentioned at the beginning of this paragraph, the  full quantum treatment is best suited when the coupling terms in the interaction Hamiltonian $H_{\rm int}$ are weak, and the model Hamiltonian $H_0$ provides already a good starting point for the coupled dynamics.

Utilizing model Hamiltonians for describing electronic dynamics is a well-established technique in the field of ultrafast soild-state optics, see e.g.~\cite{Kira99,Koch2009}. In semiconductor bulk materials and quantum wells, the dispersion $\varepsilon_{n\mathbf{k}}$ entering the Hamiltonian can be approximated as being quadratic (and sometimes as being linear, e.g. for graphene~\cite{MaWi11}), and a full solution of the relaxation dynamics for various scattering mechanisms has been achieved in these cases. 
Here, we are interested in a realistic description of the ground state of a particular system. 
For this reason, all the band energies and phonon frequencies entering $H_0$ are determined by density-functional theory calculations. 
The VASP code \cite{KrFu96} with the settings described in Ref. \onlinecite{ZaKr17} has been employed for this purpose. 
The electronic single-particle energies $\varepsilon_{n\mathbf{k}}$
are taken to be equal to the Kohn-Sham eigenvalues obtained with the
GGA-PBE exchange-correlation functional \cite{PeBu96}. 
The phonon frequencies $\Omega_{I\mathbf{Q}}$ and the corresponding eigenmodes are obtained
from DFT calculations using the method of finite atomic displacements within a supercell, as detailed
in Ref.~\onlinecite{ZaKr17,phonopy}. 
In case of the Pb/Si(111) films, such a detailed first-principles description is considered necessary in view of the experimental findings: The two-photon photoemission spectra show peaks at certain intermediate-state energies of the electrons that are referred to as quantum well states (QWS).  These are energies where the electronic density of states is high and/or where the excited electrons are long-lived. For a correct prediction of the energetic position of the QWS, the $(1 \times 1)$ periodicity of a free-standing Pb(111) films is not sufficiently accurate.\cite{SaSk14}. It is required to take the larger $(\sqrt{3} \times \sqrt{3})$ periodicity enforced by the Si(111) substrate into account. As major achievement of the first-principles calculations in Ref.~\onlinecite{ZaKr17}, we were able to reproduce quantitatively the dependence of the energetic position of the QWS on the number of Pb layers in the film, as well as the very small dispersion of the occupied QWS in the films with an odd number of Pb layers. On this basis, the present work is addressing the role of the electronic lifetime in the QWS for the experimentally detected peaks. 

The interaction Hamiltonian $H_{\rm int}$  contains any further interactions required to describe the problem at hand. 
These interactions could e.g. be the electron-electron interactions beyond the effective mean-field description of density functional theory (see below). 
Moreover, the interaction with an external electromagnetic field, e.g. of a laser pulse, can be considered as part of $H_{\rm int}$. 
Most importantly for the present study, $H_{\rm int}$ contains  a term $H_{\rm ep}$ describing in linear order 
the coupling of the electrons to quantized phonons,  
\begin{equation} \label{eq:Hint} 
H_{\rm ep} =  \sum_{ n\mathbf{k}, n'\mathbf{k}', I \mathbf{Q}}
{\bf D}_{n\mathbf{k}, I \mathbf{Q}}^{ n'\mathbf{k}' } \, 
c^{\dagger}_{n\mathbf{k}} c_{n'\mathbf{k}'} ( b^{\dagger}_{I\mathbf{-Q}} +
b_{I\mathbf{Q}})
\end{equation}
The term in parentheses is linear in each phonon coordinate. 

In principle it is possible to describe the quantum non-equilibrium dynamics under the action of $H$ exactly by a density matrix.  
Schemes for evolving the density matrix in time have been worked out \cite{RoKu02}, and applications to surfaces and low-dimensional systems can be found in the literature.\cite{Buecking2008, Richter2009, RichterMarten2009,MaWi11}
However, since the system we want to describe is quite complex, we resort to a simpler description of the dynamics which is appropriate if the coherent excitation by an optical pulse and the subsequent relaxation take place on separable time scales. 
While quantum coherence is important during the interaction of the system with the light field, electron-electron scattering usually leads to a fast loss of coherence.\cite{UeGu07} 
For Pb films, an example of calculations taking the effects of coherence into account can be found in \cite{SaMi15}. 
In the limit of vanishing coherence, only the diagonal elements of the density matrix, the populations $f_{n\mathbf{k}}$  of states indexed by $n$ and the wave vector {$\mathbf{k}$, are important.
For the investigation of the ultrafast population dynamics in our system, the quantities which we have to look at are
the electronic occupation numbers 
$f_{n\mathbf{k}} = \langle\hat{c}^\dagger_{n\mathbf{k}}\hat{c}_{n\mathbf{k}}\rangle$ 
and the phononic occupation numbers 
$n_{I\mathbf{Q}} = \langle\hat{b}^\dagger_{I\mathbf{Q}}\hat{b}_{I\mathbf{Q}}\rangle$. 
For the latter, we employ a bath approximation 
$$
{\Big\langle\hat{b}^\dagger_{I\mathbf{Q}}\hat{b}_{I\mathbf{Q}}\Big\rangle
  = n_{I\mathbf{Q}}(T_{I}) =
  \frac{1}{\exp\left( \frac{\hbar  \Omega_{I\mathbf{Q}} }{k_B T_I}\right) - 1}} \, .
$$
In the numerical calculations presented below, we will use different baths, one for each high-lying optical mode of the Pb
film  ($\Omega_{I\mathbf{Q}}  \ge 2$ THz) with 
temperature $T_I$, and a common one for all low-frequency
phonons of the Pb film ($\Omega_{I\mathbf{Q}}  < 2$ THz) with
temperature $T_0$. More details are given in the appendix.

Using the Markov approximation and the second-order Born approximation for the transitions, it is possible to derive from the density-matrix equations a set of coupled differential equations that can be cast into the form of a master equation (cf. Ref.~\onlinecite{Kuhn}). 
\begin{equation}
\frac{d}{dt}f_{n{\mathbf{k}}} = \Gamma^{\rm in}_{n{\mathbf{k}}}\left(1
  - f_{n{\mathbf{k}}}\right) - \Gamma^{\rm
  out}_{n{\mathbf{k}}}f_{n{\mathbf{k}}} \, .
\label{eq:Master} 
\end{equation}
The expressions for the rates, both for scattering into and out of the state $n\mathbf{k}$, are made up of an electronic and a phononic contribution each, i.e.,  $\Gamma_{n\mathbf{k}} = \Gamma_{n\mathbf{k}}^{(ee)} + \Gamma_{n\mathbf{k}}^{(ep)}$.  This holds for 
both  $\Gamma^{\rm in}$ and  
 $\Gamma^{\rm out}$ that both consist two terms owing to electron-electron scattering and electron-phonon scattering: 
\begin{eqnarray}
\Gamma^{\rm out}_{n\mathbf{k}} &=& \Gamma^{\rm
  out\,(ee)}_{n\mathbf{k}} + \Gamma^{\rm
  out\,(ep)}_{n\mathbf{k}} \label{eq:totRateOut} \\
\Gamma^{\rm in}_{n\mathbf{k}} &=& \Gamma^{\rm in\,(ee)}_{n\mathbf{k}}
+ \Gamma^{\rm in\,(ep)}_{n\mathbf{k}} 
\label{eq:totRateIn}
\end{eqnarray}

\smallskip

Exploiting conservation of crystal momentum parallel to the film, $\mathbf{k} - \mathbf{k'} = \pm \mathbf{Q}$ with the sign depending on phonon emission or absorption, the electron-phonon scattering rates originating from the Hamiltonian (\ref{eq:Hint}) can be expressed according to Fermi's golden rule as 

\begin{eqnarray}
\label{eq:rateEPH}
        \Gamma^{\rm in,\,(ep)}_{n\mathbf{k}} &=&\
        \frac{2\pi}{\hbar}\sum_{\substack{m,\,I,\\
            \mathbf{Q},\,\pm}}\left|{{\bf
              D}_{n\mathbf{k},\,I \mathbf{Q}}^{m\mathbf{k}-\mathbf{Q}}}\right|^2
        \left(n_{I\mathbf{Q}} + \frac{1}{2} \pm
          \frac{1}{2}\right) \times \nonumber \\
 & & \delta\big(\varepsilon_{n\mathbf{k}} -
        \varepsilon_{m\mathbf{k}-\mathbf{Q}} \pm
        \hbar\Omega_{I\mathbf{Q}} \big)
        \,f_{m,\,\mathbf{k}-\mathbf{Q}}  \\
        \Gamma^{\rm out,\,(ep)}_{n\mathbf{k}} &=&\
        \frac{2\pi}{\hbar}\sum_{\substack{m,\,I,\\
            \mathbf{Q},\,\pm}}\left|{{\bf
              D}_{n\mathbf{k},\,I \mathbf{Q}}^{m\mathbf{k}-\mathbf{Q}}}\right|^2
        \left(n_{I\mathbf{Q}} + \frac{1}{2} \mp
          \frac{1}{2}\right)  \times \nonumber \\
 &  & \delta\big(\varepsilon_{n\mathbf{k}} -
        \varepsilon_{m\mathbf{k}-\mathbf{Q}} \pm
        \hbar\Omega_{I\mathbf{Q}}
        \big)\,\big(1-f_{m,\,\mathbf{k}-\mathbf{Q}}\big) 
\end{eqnarray}

\smallskip

These expressions include processes where the electron
absorbs a phonon as well as phonon emission processes. This is denoted
by the $\pm$ signs in the equations, where the minus sign stands for
absorption, and the plus sign for both spontaneous and induced
emission, proportional to  $n_{I\mathbf{Q}} + 1$.

\begin{figure*}[tbp]
\begin{center}
\includegraphics[width=0.9\textwidth]{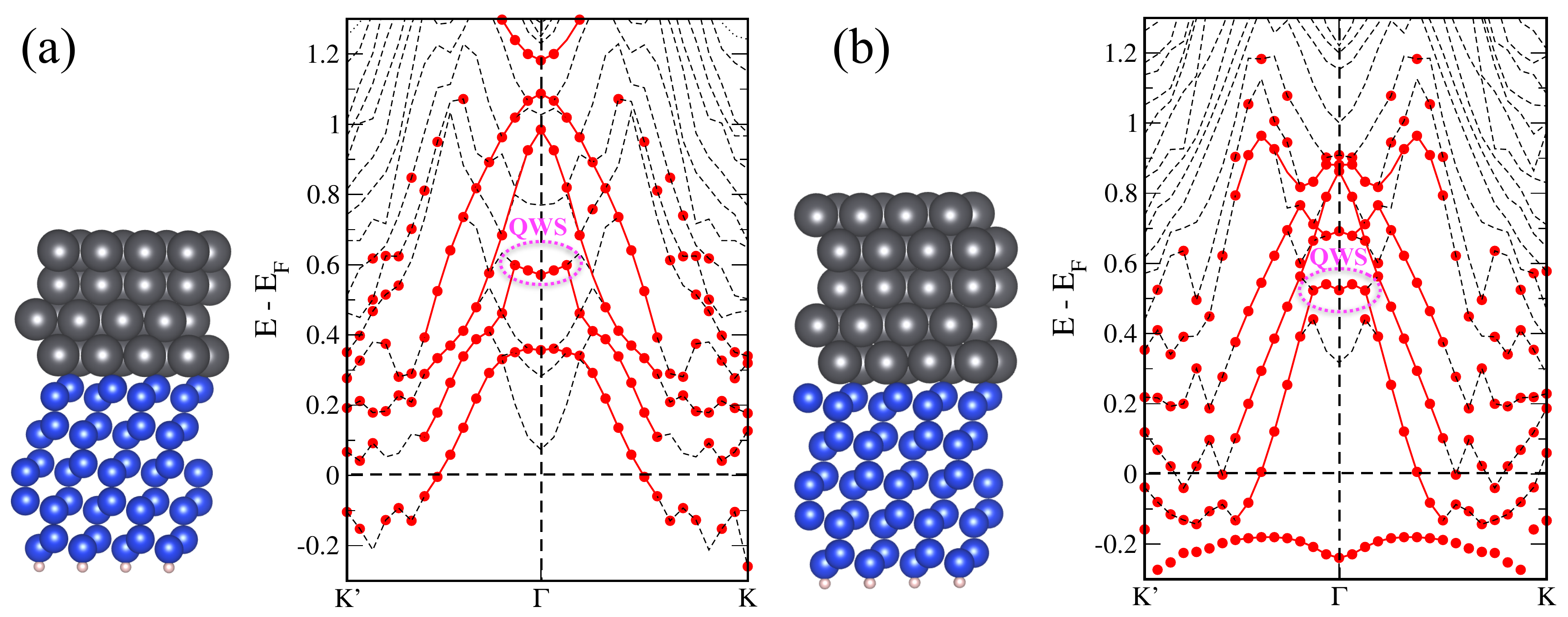}
\caption{Atomic structure and electronic bands (dashed lines) in a selected energy range above the Fermi energy (horizontal dashed line) for (a) 4~ML Pb on Si(111)$(\sqrt{3} \times \sqrt{3})$ (b) 5~ML Pb on Si(111)$(\sqrt{3} \times \sqrt{3})$.  
The bands with large Pb $6p_z$ character are highlighted by the red symbols and thick red lines.  
The dashed ellipses mark the regions of quantum wells states (QWS) whose lifetime under e-ph scattering is shown in Fig.~2(a).
For 5~ML Pb, the occupied quantum well resonance at $\sim -0.25$ eV has been included in the plot. 
The electronic eigenvalues are represented on a $32 \times 32$ Monkhorst-Pack grid. 
The plots are shown along a diagonal cut K' - $\Gamma$ - K through the Brillouin zone of the $(\sqrt{3} \times \sqrt{3})$ cell.
\label{fig:PbBandst}
}
\end{center}
\end{figure*}


It is our goal to calculate the contribution of e-ph scattering
to the lifetime of specific quantum well states (QWS) in Pb/Si(111) films. 
In a very simple picture, the conduction band electrons of Pb with crystal momentum normal to the surface or interface of the Pb(111) films are confined, similar to the quantum-mechanical particle-in-a-box problem. 
In an atomistic picture,  these conduction band states are derived from the $6p_z$ orbitals of the Pb atoms and their wavefunctions extend both above the surface and into the Si(111) substrate, see Ref. \cite{ZaKr17}.
As described in the experimental paper \cite{Kirchmann2010}, there are significant differences between the lifetimes in films with an even and an odd number of Pb layers.  Therefore, we study two representative systems, a Pb film with 4 monolayers (ML) and one with 5~ML on Si(111). 
Side views of the corresponding slabs are depicted in Fig.~\ref{fig:PbBandst}. 
Motivated by the experimental focus on excited electrons in unoccupied bands, we include e-ph scattering
rates for the {\em electrons} excited into QWS. Since the population of the valence bands was not analyzed in these experiments, 
the hole states are treated in less detail, and 
and only Coulomb scattering, as described in Section~\ref{sec:ee}, will be considered among the holes.  
To solve the rate equations, we need explicit expressions for the  quantities
$D_{n\mathbf{k},\, I \mathbf{Q}}^{ m\mathbf{k}-\mathbf{Q} }$ and
$n_{I\mathbf{Q}}(T)$ in eq.~(\ref{eq:rateEPH}) entering the decay rates $\Gamma^{{\rm in, \,
  (ep)}}$ and $\Gamma^{{\rm out, \, (ep)}}$.  
Both  quantities depend on the phonon branches $\Omega_{I\mathbf{Q}}$.  
Of all phonon modes $\Omega_{I\mathbf{Q}}$ of the supercell obtained with our first-principles approach \cite{phonopy}, those with Pb
character are taken into account, see Fig.~6 in Ref.~\onlinecite{ZaKr17}. This amounts to 
$I = 1, \ldots 48$ for the 4~ML Pb slab and $I =1, \ldots 60$ for the 5~ML Pb film on Si(111). 
To keep the number of individual scattering processes at a tractable level,
we also restrict ourselves to a subspace of the electronic bands: Since we are interested
in the electron-phonon coupling in QWS in Pb, only 
those electronic bands that have a significant overlap with the Pb $6p_z$
orbitals, as indicated by the VASP calculation, are retained in the 
Hamiltonian $H_{\rm ep}$ in eq.~(\ref{eq:Hint}). 
The electronic states belonging to a specific Pb-derived band are
grouped together into subsets indexed by $\alpha(\mathbf{k}) \in  \{n\}$ of all band
indices $n$.  To be specific, we used the five (six) lowest-lying conduction bands with appreciable Pb
$6p_z$ character for the 4~ML and 5~ML Pb film, respectively, i.e. $\alpha =1,\ldots 5, (6)$. 
These 5 (6) bands are displayed  in Fig.~\ref{fig:PbBandst} by the thick red lines and symbols, 
together with the full band structure (dashed lines)  that is also shown (over a wider range of energies and  wavevectors) in Fig.~3 and 4 of Ref.~\onlinecite{ZaKr17}. 
Due to the use of a supercell and backfolding of the bands, these bandstructures are different from the bandstructure of Pb(111)$(1 \times 1)$ slabs that had  been used previously\cite{Kirchmann2010} in the experimental data analysis. 

For evaluating the electron-phonon scattering rates, eq.~(\ref{eq:rateEPH}), we use techniques based on deformation potential theory that 
allows us to obtain  $\Gamma^{\rm out\,(ep)}_{n\mathbf{k}}$ from first-principles calculations of the phonon spectrum, the 
electronic wavefunctions, and Kohn-Sham eigenvalues, with only few approximations.  
As the most significant one, we neglect of the $\mathbf{Q}$-dependence of the deformation potential, 
while keeping its dependence on band index $n$ and crystal momentum $\mathbf{k}$. 
This is a good approximation for optical phonons and corresponds to
keeping the leading (constant) term in an expansion in powers of
$\mathbf{Q}$, cf. Ref.~\onlinecite{Resta}. 
In the energy-conserving $\delta$-function in eq.~(\ref{eq:rateEPH}), we  retain the finite
phonon energy $\hbar  \Omega_{I\mathbf{Q}} \approx \hbar  \Omega_{I 0}$, but neglect the dispersion of the optical phonon branches. 
This is justified since the dispersion remains small (cf. Fig.~6 in Ref.~\onlinecite{ZaKr17}) due to the large real-space unit cell, and hence 
small Brillouin zone, of the Pb films. 
Within these approximations 
(see appendix for the derivation), the matrix element for electron-phonon scattering in eq.~(\ref{eq:rateEPH}) can be replaced by 
\begin{equation}
 \label{eq:Dnm}
|{\bf D}_{n\mathbf{k},\,I\mathbf{Q}}^{ m\mathbf{k}' }|^2 \approx 
\frac{\hbar |D_{n\mathbf{k}, \, I}|^2 }{2\Omega_{I 0}M_{\rm
    Pb}}\frac{v_{\mathrm{atom}}^{\mathrm{Pb}}}{A_{\mathrm{supercell}}}\,I_{n\,\mathbf{k}}^{m\mathbf{k}'}\,\delta_{\mathbf{k-k',Q}}
\, \delta_{m,\alpha(\mathbf{k}')}
\end{equation}
Here $A_{\mathrm{supercell}}$ is the area of the  Si(111)$(\sqrt{3} \times \sqrt{3})$ supercell used to
model the Pb/Si(111) film, $M_{\rm Pb}$ and $v_{\mathrm{atom}}^{\mathrm{Pb}}$ 
are the atomic mass and atomic volume of Pb.   
$D_{n\mathbf{k}, \, I}$ is the deformation potential of the $n^{\mathrm{th}}$
electronic band under the phonon mode $I$.  
The $D_{n\mathbf{k}, \, I}$ have been obtained from DFT calculations
\cite{ZaKr17} by evaluating the electronic eigenvalue shift
under finite displacements of the atomic positions given by the
corresponding mode eigenvector of the phonon. 
The two $\delta$-symbols reflect conservation of crystal momentum in e-ph
scattering, and the projection of $H_{\rm ep}$ to the finite
electronic subspace, as described above. 
The matrix elements $I_{\,n\mathbf{k}}^{m\mathbf{k}'}$ account for
the difference between intra-band ($n=m$) and interband scattering ($n \ne m$),
and for the dependence on {\em both} the initial and final electron momenta $\mathbf{k}$
and $\mathbf{k'} = \mathbf{k} - \mathbf{Q}$. They are obtained from the overlap of the
corresponding DFT wave functions. 
More details of the derivation are given in the appendix.  
In summary, this approach allows us to arrive at a simplified and computationally tractable, yet parameter-free
description of e-ph scattering even for such a complex systems as an
overlayer on a substrate. 

\begin{figure}[tbp]
\begin{center}
\begin{tabular}{lc}
(a) & \\
& \includegraphics[width=0.3\textwidth]{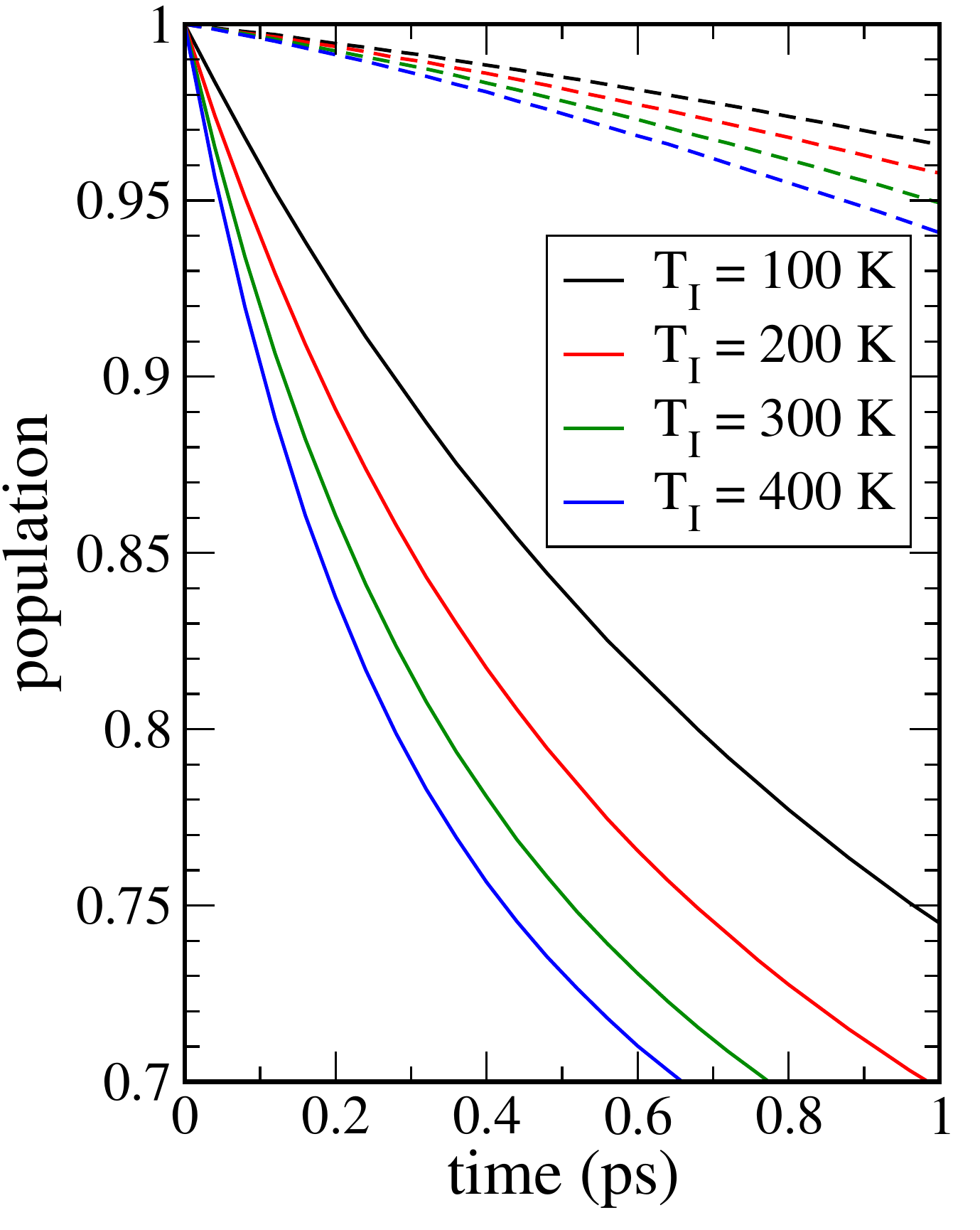} \\ 
(b)&  \\ 
& \includegraphics[width=0.3\textwidth]{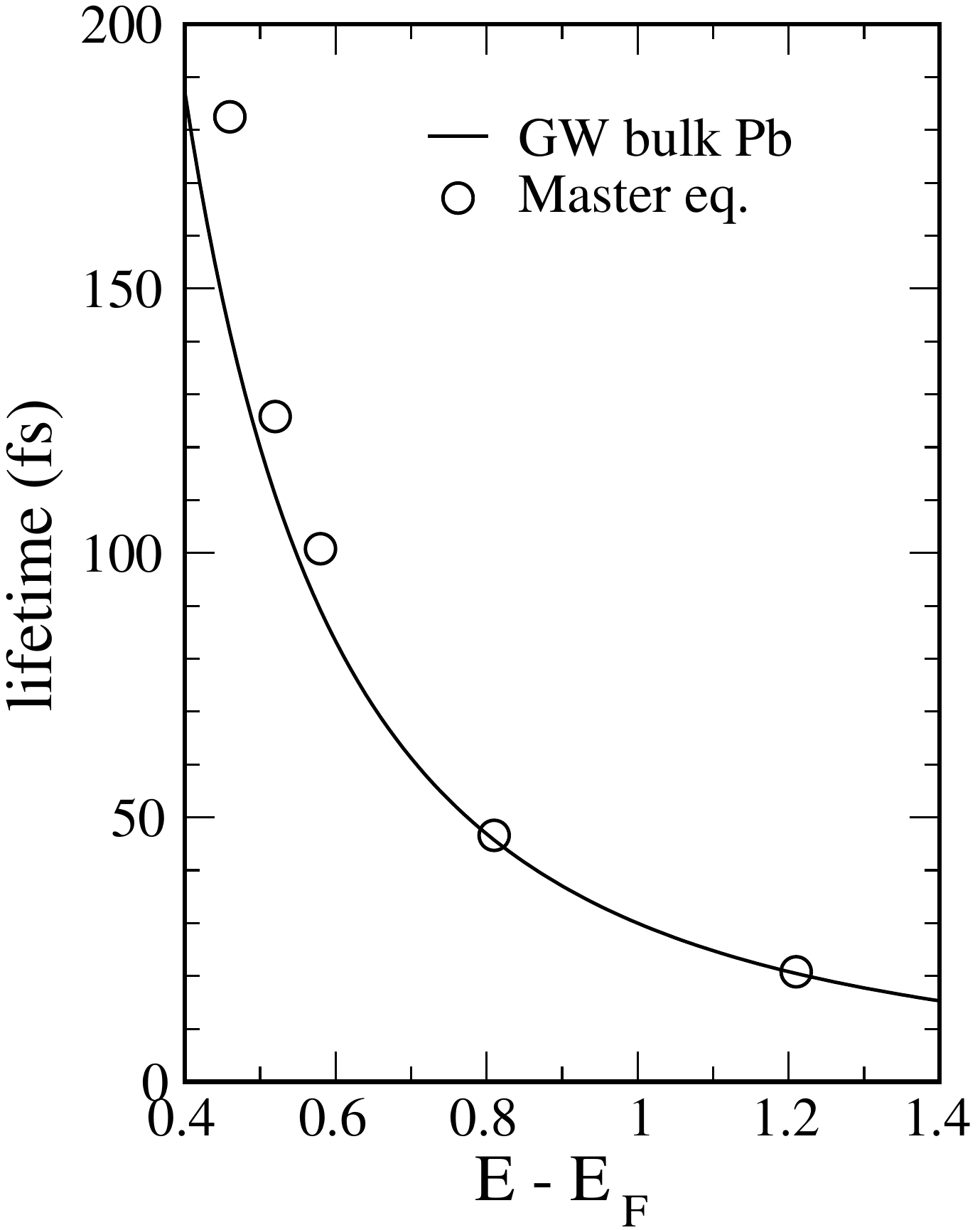}
\end{tabular}
\caption{a) Decay of the population in a single quantum well state at the $\Gamma$-point, as indicated by the ellipses in Fig.~1, solely due to e-ph scattering. 
The energy of the chosen initial state was 0.58~eV for the 4~ML Pb film (full lines) and 0.53~eV for the 5~ML Pb film (dashed lines). 
Results for various temperatures $T_I$ of the phonon heat bath between 100~K and 400~K are shown. 
b) Lifetime of excited electrons due to e-e scattering, eq.~(\ref{eq:fermiliquid}) (full line) as a function of energy . 
The circles show the lifetime of quantum well states in a 4~ML and a 5~ML Pb film as obtained from solving the master equation including both e-e and e-ph scattering.}
\label{fig:lifetime}
\end{center}
\end{figure}     

\section{Results}

In metals, e-ph and e-e scattering are closely intertwined, since the vast majority of phonons is emitted by secondary electrons and holes rather than by the charge carriers initially excited by the light pulse. This is because e-e scattering quickly generates an avalanche of secondary electron-hole pairs with small energies around the Fermi level. Since these secondary electrons and holes are produced with high density and  their energy still exceeds typical phonon energies, they play a major role in determining the rate at which the energy is dissipated from the electronic system into the lattice. 
Nevertheless, we start our discussion by considering the contribution of both e-ph and e-e scattering separately.

\subsection{Relaxation due to e-ph scattering}

First we investigate how the population of a QWS decays under the sole effect of e-ph scattering.  
For this purpose, we initially populate a single QWS at the $\Gamma$-point and let the population evolve according to the master equation (\ref{eq:Master}) using only the rates  $\Gamma^{{\rm in, \,
		(ep)}}$ and $\Gamma^{{\rm out, \, (ep)}}$.  
The results are shown in Fig.~\ref{fig:lifetime}(a). 
At comparable energies of the QWS of $\sim 0.5$eV, the decay is much faster in the 4~ML than in the 5~ML Pb film. 
This is to be expected from the different size of the deformation potentials in the two films reported in Ref.~\cite{ZaKr17}. 
The relaxation rate increases with the temperature of the phonon heat bath, which is indicative of the role of stimulated emission of phonons. 
By decreasing the phonon temperature from 400~K and 100~K, the lifetime of the QWS in the 4~ML film increases from 
1.3 to 2.7~ps. For the 5~ML film, the lifetimes fall between 13 and 37~ps.

\subsection{Relaxation due to e-e scattering \label{sec:ee} }

The lifetime of hot electrons due to e-e scattering can be described by a
 self-energy formalism, as discussed in Ref.~\onlinecite{LaHo04}. 
The loss term $\Gamma^{\rm out\,(ee)}_{n\mathbf{k}}$ in eq.~(\ref{eq:totRateOut}) is given by
$\Gamma^{\rm out,\,(ee)}_{n\mathbf{k}} = -2{\rm Im} \Sigma (\varepsilon_{n\mathbf{k}})/\hbar$.  
The self-energy $\Sigma$ is obtained from a $GW$ calculation of bulk Pb.
Here, $G$ stands for the electronic Green function, and $W$ for the screened Coulomb interaction. 
These quantities are calculated from the DFT wave functions and Kohn-Sham eigenvalues using the built-in capabilities of VASP \cite{ShKr06}. 
To be specific, a $11 \times 11 \times 11$ k-point mesh is used, and the denominator in $G$ is evaluated with a small shift of the transition energy away from the real axis, $\eta = 0.08$~eV, much smaller than typical values used in $GW$ calculations of band structures. 
The result obtained for $- 2 {\rm Im} \Sigma$ in the conduction band is fitted
to the $\alpha (\varepsilon_{n\mathbf{k}} - E_F)^2$ dependence expected from Landau's theory of the Fermi liquid. 
Our result $\alpha = 0.022$ (eV)$^{-1}$ is in excellent agreement with
earlier $GW$ calculations of bulk Pb \cite{HoBr09}. Finally, we obtain the
expression 
\begin{equation}
        \Gamma^{\rm out,\,(ee)}_{n\mathbf{k}} =
        \frac{(\varepsilon_{n\mathbf{k}} - E_{F})^2}{30 \mathrm{fs
            (eV)}^2} 
            \label{eq:fermiliquid}
\end{equation}
plotted in Fig.~\ref{fig:lifetime}(b).

In the relaxation of highly excited electrons and holes, the energy is dissipated to secondary electron-hole pairs. This process is very efficient in metals, since, in contrast to semiconductors, there is no energy gap preventing the generation of secondary particles. These effects are included in the 
the scattering-in term $\Gamma^{\rm  in\,(ee)}_{n\mathbf{k}}$  of eq.~(\ref{eq:totRateIn}). 
Although this term can be obtained
from the master equation \cite{LaHo04} as well, we choose for computational convenience a
simpler treatment in our present study. The gain term is assumed to
factorize into an energy-dependent and a time-dependent factor, 
$
        \Gamma^{\rm in,\,(ee)}_{n\mathbf{k}} = \Phi(x) N(t).
$
The  distribution function $\Phi$ describes
the secondary electrons and holes produced via impact ionization by a relaxing high-energy electron.
Following the work of Baranov and Kabanov~\cite{BaKa14}, we use 
for $\Phi$ a stationary solution of the Boltzmann equation with a Coulomb scattering kernel, 
$\Phi(x) = x/\cosh^2 (x/2), \; x=(\varepsilon_{n\mathbf{k}} - E_{F})/(k_{B}T_{\rm el})$,  
where the electronic temperature $T_{\rm el}= 650$~K 
was chosen in accordance with the energy of $E_{\rm dep} = 0.1 {\rm eV} = \frac{\pi^2}{6} g(E_F) (k_B T_{\rm el})^2$ deposited
by the laser and the electronic heat capacity of Pb $c_V = \frac{\pi^2}{3} g(E_F) k_B^2 T_{\rm el}$. 
The time-dependent factor $N(t)$ for creation of secondary electrons
and holes is determined by energy conservation in the e-e scattering.  
Our simplified treatment assumes that the initial electron in state $n\mathbf{k}$ 
 ends up at the Fermi energy, transferring all its initial energy to secondary
 electron-hole pairs. This motivates the choice 
\begin{equation*}
N(t) = \frac{\sum_{n,\mathbf{k}} |\varepsilon_{n\mathbf{k}}-E_{F}| \, \Gamma^{\rm out, \, (ee)}_{n\mathbf{k}} \, f_{n\mathbf{k}}}{\int d \varepsilon g(\varepsilon) \, |\varepsilon -E_{F}| \, \Phi((\varepsilon - E_{F})/k_BT_{\rm el}) }    
\end{equation*}
where $g(\varepsilon)$ is the electronic density of states.  
Both $g(\varepsilon)$ and $N(t)$ are evaluated numerically using as input the
DFT band structure of slab models for Pb/Si(111)$(\sqrt 3 \times \sqrt 3)$ multilayer films. 

\begin{figure}[tbp]
\begin{tabular}{lc}
(a)  & \\
& \includegraphics[width=0.5\textwidth]{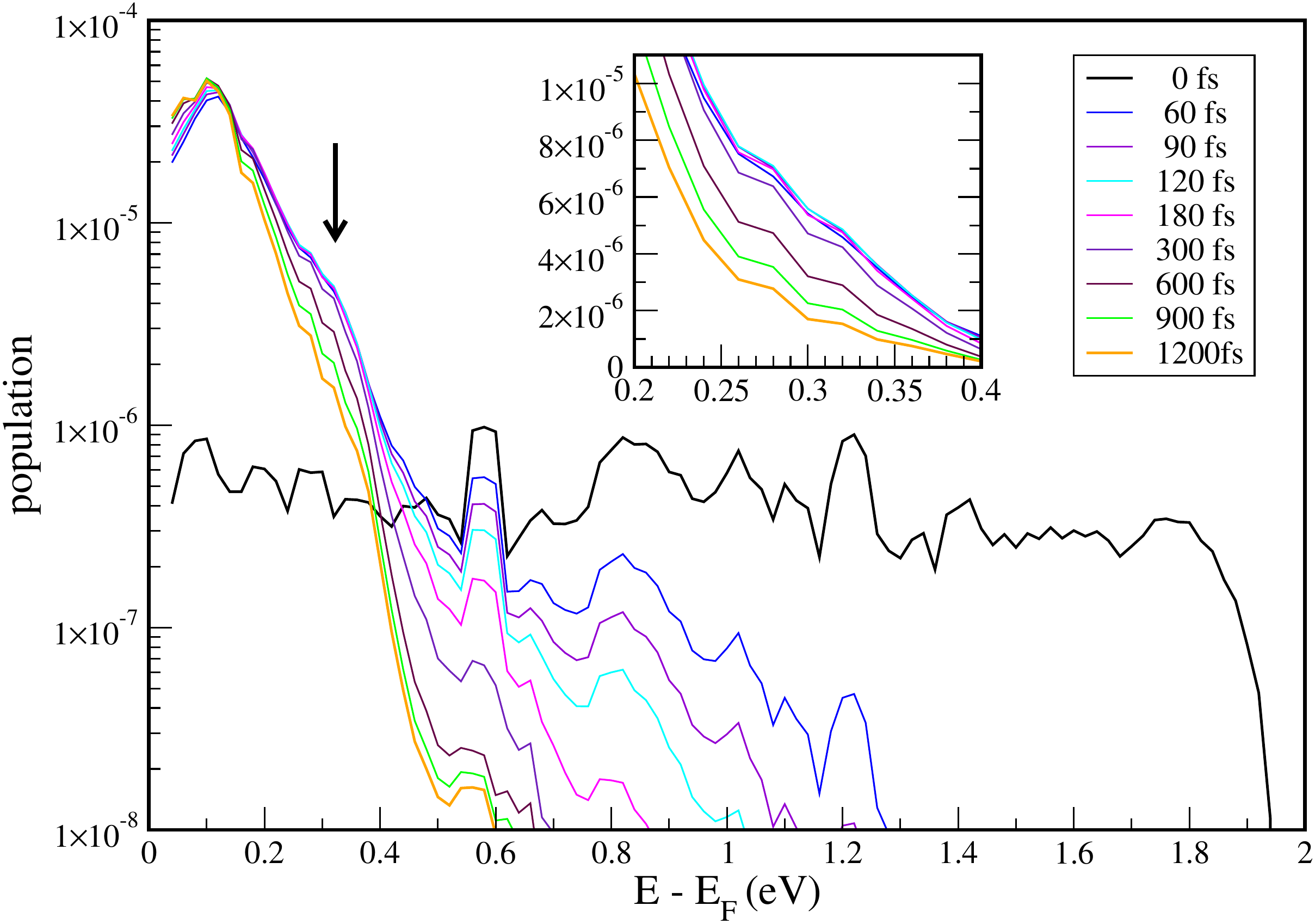} \\
(b)  & \\
& \includegraphics[width=0.5\textwidth]{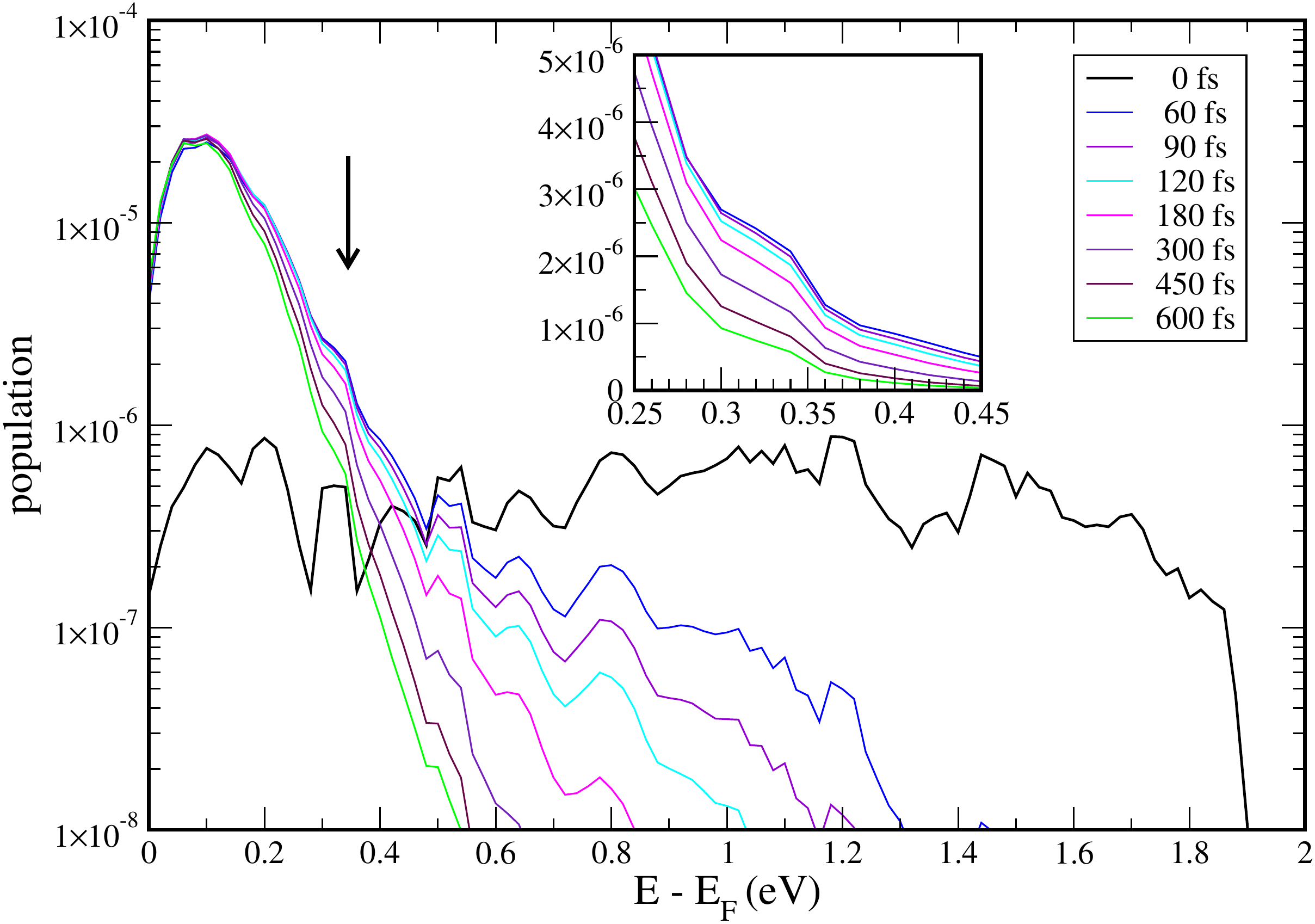}
\end{tabular}
\caption{Electronic excitation spectra  (thick black line) after a laser pulse with $h \nu =1.9$ eV for (a) 4~ML and (b) 5~ML of Pb/Si(111).  
The colored lines show the population of electronic states after a time delay of 60, 90, 120, 180, 300, 600, 900 and 1200~fs after excitation. 
The approximately exponential part of the spectra at low energies results from secondary electrons excited via e-e scattering. The arrows mark the energy range where, in addition to the secondary-electron distribution,  a delayed relaxation of hot electrons due to a 'phonon bottleneck' can be observed. A magnification of this region is shown in the inset.}
\label{fig:enedist}
\end{figure}        

\subsection{Competition between e-e and e-ph scattering}

In this Section, we compare simulation results for Pb films of 4~ML and 5~ML thickness as representatives of films with an even and odd number of layers studied by optical pump-probe experiments in  Ref.~\onlinecite{Kirchmann2010}. 
While these experiments measure the total probability for two-photon photoemission, our simulations model the population of the intermediate electronic states that are reached by the electron after applying the pump pulse and subsequent relaxation. The second step of the two-photon photoemission, which kicks the electron into the vacuum, is not modeled.  Provided that the probability of ionization by the probe pulse is a smooth function of energy, the measured yield can be considered approximately proportional to the population of the intermediate state.

The initial distribution is chosen such that it describes the response of our specific system, multilayers of Pb on Si(111), to a short optical pulse with frequency centered around $h\nu=1.9$~eV. This corresponds to the photon energy of the pump laser used in the experiment \cite{Kirchmann2010}. 
The polarization of the electric field, denoted by the unit vector
$\mathbf{e}$,
is chosen parallel to the Pb film surface.  
Before the laser pulse arrives, the system is described by a Fermi-Dirac distribution with low temperature, $T_{\rm el} \to 0$; hence 
$f_{n{\mathbf{k}}}(t<0) =   \Theta( E_F - \varepsilon_{n \mathbf{k}} )$. 
To be specific, we evaluate dipole matrix elements\cite{GaHu06} 
\begin{eqnarray}
f_{n{\mathbf{k}}}(t=0) &=& A_0 \sum_{m}  \bigl| \langle n \mathbf{k} | \mathbf{e} \cdot \nabla  | m \mathbf{k}  \rangle \bigr|^2   \delta( \varepsilon_{n \mathbf{k}} - \varepsilon_{m \mathbf{k}} - h \nu) \times \nonumber \\
& & \bigl( \Theta(E_F - \varepsilon_{m \mathbf{k}} )  - \Theta( E_F - \varepsilon_{n \mathbf{k}} ) \bigr)
\label{eq:dipole}
\end{eqnarray}
In the numerical evaluation
\footnote{The dipole matrix elements are delivered by the VASP code using the keyword LOPTICS}, a broadening of the $\delta$-function by 0.02~eV is used. 
The proportionality factor $A_0$ is chosen 
such that the  energy of excited electrons and holes deposited in the Pb films amounts to $\sim 0.1$~eV per supercell area, 
equivalent to 3.7~$\mu$J/cm$^2$. 

By solving the master eq.~(\ref{eq:Master}) numerically, we are able to follow the relaxation of the excited electrons in real time. We define an energy and time dependent population 
of the intermediate state 
\begin{equation}
P(\varepsilon, t) = \sum_{n, \mathbf{k}} f_{n,\mathbf{k}}(t) \delta(\varepsilon - \varepsilon_{n, \mathbf{k}}) \, .
\label{eq:energyDist}
\end{equation}
Fig.~\ref{fig:enedist} shows on a logarithmic scale the energy  distribution $P(\varepsilon, t_j)$ of the excited electrons for various times $t_j$ after the excitation. 
For plotting the results, the $\delta$-function in eq.~(\ref{eq:energyDist}) has been replaced by a rectangle with a  width of 0.06~eV.

We start with a discussion of the initial distribution, shown by the thick black line, calculated according to the transition dipole  strength, eq.~(\ref{eq:dipole}). For the 4~ML Pb film (Fig.~\ref{fig:enedist}(a)), the distribution is highly structured with a sharp maximum at 0.58~eV and a broad peak around 1.21~eV.  
For the 5~ML Pb film (thick black line in Fig.~\ref{fig:enedist}(b)), only the peak at 1.21~eV (and possibly a short-lived peak at higher energies) remain visible, while the low-energy peak is much less pronounced.  
These results are in excellent agreement with the experimental observations of  Ref.~\onlinecite{Kirchmann2010}. In this work, a high-energy peak in the range of 1.1 to 1.2~eV was observed for films with an odd number of Pb monolayers, 
whereas the peak at 0.6~eV was dominant in Pb films with an even number of layers. 
Note that, due to experimental limitations of the probe laser energy, excited electrons with energies lower than $\sim 0.5$~eV could not be detected in Ref.~\onlinecite{Kirchmann2010}.

Next, we analyze the relaxation of the energy distributions 
for later times. From Fig.~\ref{fig:enedist} it is obvious that all distributions develop a low-energy part corresponding a quasi-thermal distribution of secondary electrons, showing up as an exponentially decreasing  function of energy. 
The high-energy part of the initial spectrum decays mainly due to e-e scattering, thereby creating secondary  electrons via impact ionization. Therefore, the high-energy tails decay quickly, simultaneously accompanied by an increasing weight of the secondary-electron distribution.
Now turning to longer time scales, we observe that 
the low-energy part in the 4~ML Pb film and the population in the energy window between 0.4 and 0.6~eV decay more slowly. At the same time, a broad shoulder on top of the secondary electron distribution builds up at $\sim 0.3$~eV (marked by the arrow in Fig.~\ref{fig:enedist}(a) and magnified in the inset). Both phenomena can be traced back to the effect of e-ph scattering: Excited electrons of the initial distribution at energies of 0.4 to 0.6~eV 'glide down' the electronic band structure (cf. Fig.~\ref{fig:PbBandst}(a) ), thereby emitting phonons. The relatively slow rate at which this occurs leads to a 'phonon bottleneck', i.e., to the build-up of the shoulder centered at $\sim 0.3$~eV. 
This effect, related to the discreteness of electronic states, is quite common for the e-ph relaxation in nanostructures, and has been observed e.g. in quantum dots \cite{Heitz2001} as well as in two-dimensional layered semiconductors \cite{Jia2018}.
While there is still continuous energy dissipation to the crystal lattice by very low-energy secondary electrons, the 'phonon bottleneck' affects electrons at higher energies and results in additional, but delayed production of phonons emitted by these electrons 'gliding down' the conduction bands. 
A similar, but weaker 'phonon bottleneck' can be observed on top of the secondary electron distribution in the simulations for the 5~ML Pb film in Fig.~\ref{fig:enedist}(b), marked by the arrow and magnified in the inset. 

\begin{figure}[tbp]
\begin{tabular}{lc}
(a) & \\ 
& \includegraphics[width=0.4\textwidth]{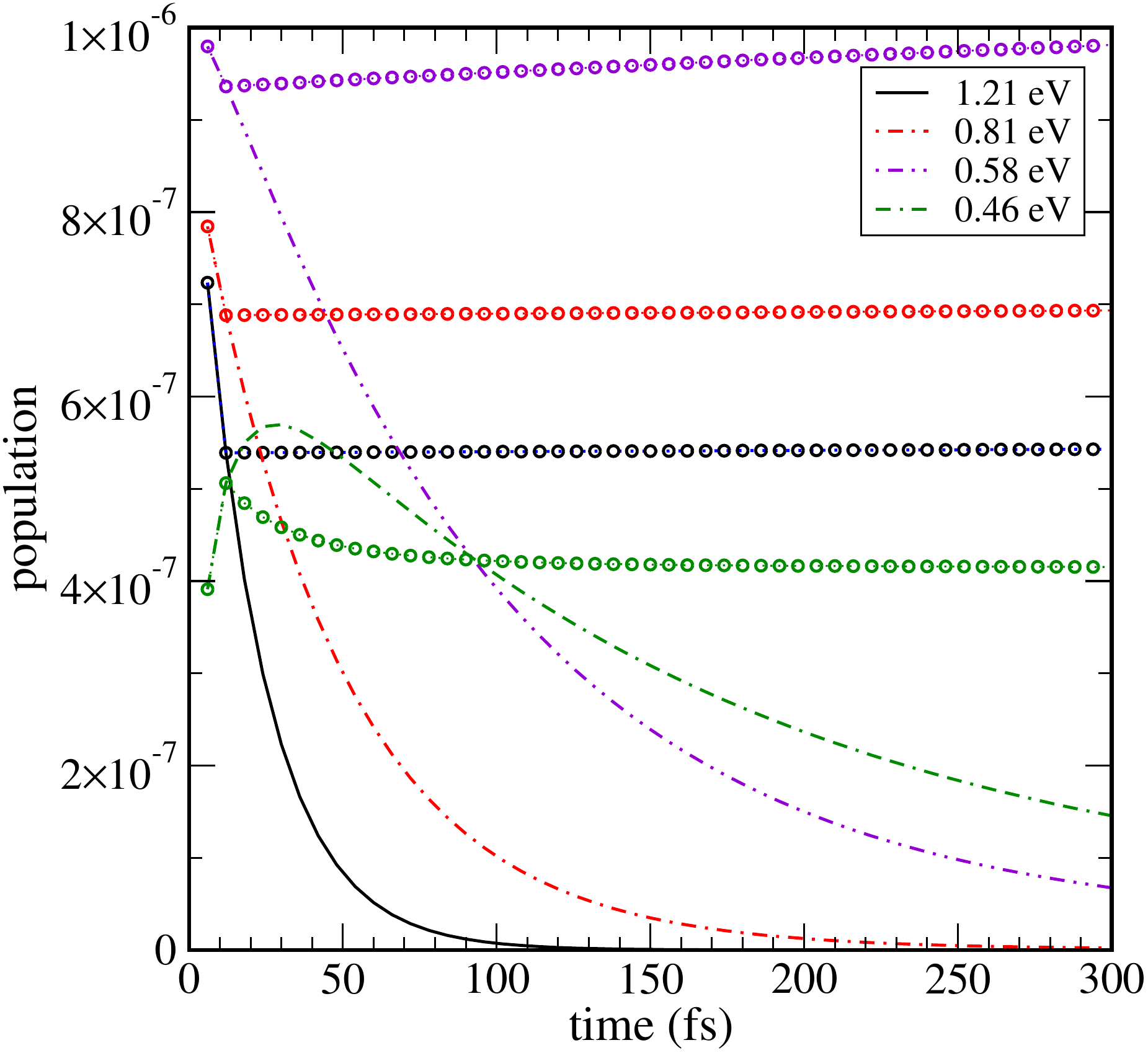} \\ 
(b) &  \\
& \includegraphics[width=0.4\textwidth]{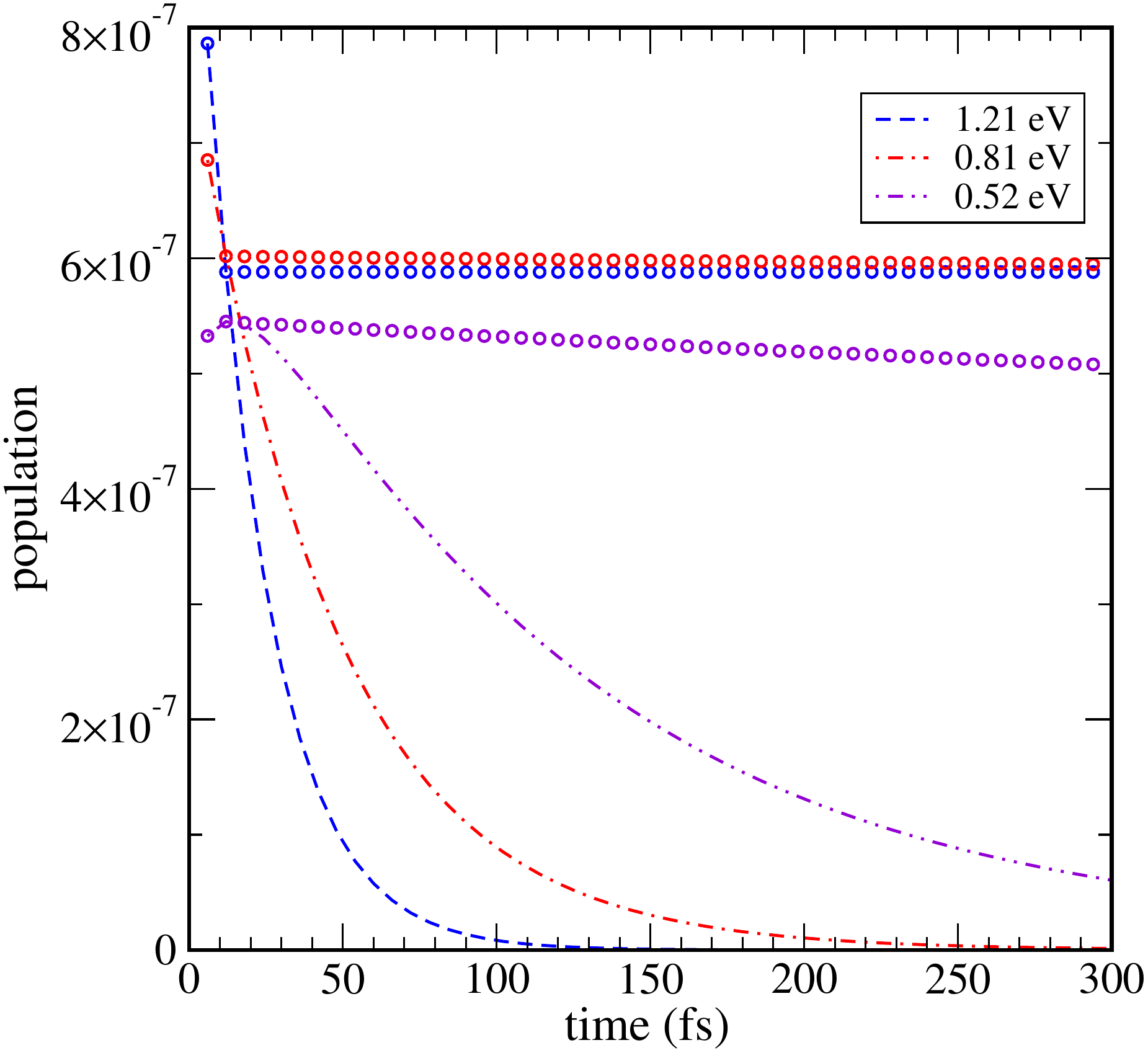}
\end{tabular}
\caption{Relaxation of the population at various electronic energies $E-E_F$ for (a) 4~ML and (b) 5~ML of Pb/Si(111). The lines give the relaxation under both e-e and e-ph scattering. The lines marked with circles show the relaxation dynamics if e-e scattering is turned off after 6~fs, and the population evolves solely under e-ph scattering.}
\label{fig:decay}
\end{figure}        

In Fig.~\ref{fig:decay}, we analyze the time scales associated with the electronic population decay at selected  energies where peaks had been found in the populations in Fig.~\ref{fig:enedist}. 
The lines without symbols in Fig.~\ref{fig:decay} show the decay according to the full relaxation dynamics, including both e-e and e-ph scattering. 
All populations show a nearly exponential decay, albeit with different decay rates. 
The lifetimes for the population maxima  at 1.21~eV and 0.58~eV, extracted via exponential fits, are 
21~fs and 101~fs, respectively. 
For the lowest energy of 0.46~eV, we find an initial rise of the population due to scattering-in from electrons at higher energies, followed by a population decay after about 30~fs, corresponding to a lifetime of 183~fs. 
The rather broad (in time) maximum of the 0.46~eV curve results from a compensation of the rates of incoming electrons from higher energies, mostly originating from e-e scattering at these high energies, and losses due to both e-e and e-ph scattering, the latter one gaining in relative importance as we go to lower energies. Interestingly, the described broad maximum of the population evolution at low energies is also seen in the experimental data \cite{Kirchmann2010}.
In the computer simulation, we can deliberately turn off the e-e scattering channel after a very short initial time interval.  
The e-e scattering was permitted only in the very early times, 
somewhat arbitrarily chosen to be less than 6~fs, since some mechanism is required to establish a realistic smoothened electron distribution including an appropriate low-energy secondary-electron part.  
The result of these runs are displayed in Fig.~\ref{fig:decay} by the lines with circular symbols. 
If the relaxation after 6~fs proceeds by e-ph scattering only, the population at 0.46~eV initially decays on a time scale of 350~fs (green symbols in Fig.~\ref{fig:decay}(a)), 
which can be taken as an estimate of the e-ph scattering rate in this energy range. This initial decay is followed by a much slower decay over several picoseconds. 
At the higher electron energies of 0.58, 0.81 and 1.21 eV, the scattering-in events of electrons from higher energies are equally probable or even more frequent than the scattering-out events, and hence a net  contribution of e-ph scattering to the decay is not detectable on the time scale shown in Fig.~\ref{fig:decay}. 

A similar analysis has been carried out for the 5~ML Pb film, see Fig.~\ref{fig:decay}(b). For the peak energies at 1.21, 0.81 and 0.52~eV, overall lifetimes of  
21, 47 and 126~fs are obtained from exponential fits to the full relaxation dynamics. Again, it is  possible to estimate the relative importance of e-ph scattering by watching the population decay after the e-e scattering has been 'turned off'. From the slopes of the curves marked by the circular symbols in Fig.~\ref{fig:decay}(b), characteristic times of 24~ps and 4.1~ps are obtained for 0.81 and 0.52~eV electron energy, respectively. 
At the highest electron energy of 1.21~eV, again we find that the contribution of e-ph scattering is too small to be detectable on the time scale shown in Fig.~\ref{fig:decay}. 

From this analysis, we learn that the contribution of e-ph scattering to the total lifetime of the peaks at energies larger than 0.5~eV is much smaller compared to the e-e contribution. This finding confirms the original analysis of the experimental data by Kirchmann {\em et al.}~\cite{Kirchmann2010} where e-ph scattering had been disregarded. Analyzing the experimental data for many Pb film thicknesses, they concluded that the low-energy peak is clearly observed in films with an even number of atomic layers and has a lifetime of $115 \pm 10$~fs, while the high-energy peak is visible only in the odd-layer films and has an energy-dependent lifetime which turns out to be $10\pm 5$~fs for 5~ML Pb. 
Our simulation results of 101~fs and 21~fs are in reasonable agreement with their experimental findings, in particular if it is taken into account that the e-e scattering rate is very sensitive to the precise energetic position of the peak. 
The lifetimes extracted from our simulations are summarized by the circular symbols in Fig.~\ref{fig:lifetime}(b). 
Despite the additional decay channel of e-ph scattering being taken into account, the simulated lifetimes lie above the lifetime of isolated electrons due to e-e scattering alone. This is because the simulations describe a realistic distribution of excited electrons, and the incoming flux from higher-lying electronic states 
effectively 'conserves' the population of the lower lying states over longer times.  
 
\subsection{Excitation of lattice vibrations}

Although the contribution of e-ph scattering to the lifetime of the quantum well states was found to be small, the low-energy states populated by the secondary electrons couple significantly to the lattice vibrations.  
At these low energies, the e-ph scattering as loss mechanism even dominates over e-e scattering, since the lifetime due to e-e interactions rises above 300~fs for electrons below 0.33~eV according to eq.~(\ref{eq:fermiliquid}). 
With the help of the simulations, it is possible to follow the energy transferred from the electrons to each of the phonon modes separately.  
Fig.~\ref{fig:esume} shows the increase in time of the excess vibrational energy (in addition to the thermal energy corresponding to the initial substrate temperature) in the various Pb vibrational modes. 
Summation over all modes yields a total energy transfer between the electronic and the lattice degrees of freedom of 8.2~meV/ps for the 4~ML Pb and 0.79~meV/ps the 5~ML Pb film, respectively. 
The smallness of these quantities (a few meV compared to 0.1 eV electronic energy in the film) gives an {\em a posteriori} justification for the perturbative expression used for the interaction Hamiltonian $H_{\rm int}$. 
It is seen from Fig.~\ref{fig:esume} that the energy transfer is highly mode-selective\footnote{In our simulations, we explicitly allowed for the possibility of conversion between different vibrational modes (see appendix), but it should be noted that vibrational coupling takes place on a much longer time scale of at least 30~ps and is hardly relevant for the observations in Fig.~\ref{fig:esume}.}. 
One particular surface phonon mode receives a major part of the energy. 
The dominance of this single mode depends on film thickness; in the 4~ML film it is clearly more pronounced than in the 5~ML film where several modes participate in the energy uptake.
In the 4~ML Pb film, this is a mode with frequency 2.26~THz (labeled 47 in our previous publication \cite{ZaKr17}), while in the 5~ML Pb film it is a phonon mode at 2.03~THz. 
In experiments using a high laser fluence, the strong coupling to a specific mode 
has indeed been observed; 
in the 5~ML Pb film, it was detected experimentally by a periodic shift of the quantum well energy with a frequency of $2.0 \pm 0.1$~THz 
\cite{ReKi12}, which is in excellent agreement with the frequency of 2.03~THz identified in our simulation. 
For both film thicknesses, the modes at frequencies below 2~THz, which are similar to phonon modes in bulk Pb, receive only a much smaller amount of energy on average.  

As seen from Fig.~\ref{fig:esume}, the increase of the energy over long time scales, e.g. over 4~ps, is sub-linear and saturation is expected at even longer times. We attribute this slow response to the rather long time required by the electrons to relax from high energies down to $E_F$ via multiple phonon emission processes.  Due to the 'phonon bottleneck' described above, this takes longer than expected from Allan's formula\cite{Allan87}. This formula requires as sole input the electron-phonon coupling constant of the bulk material that can e.g. be determined from ultrafast reflectivity measurements~\cite{BrKa90}.  

\begin{figure}[tbp]
\begin{tabular}{lc}
(a) &  \\ 
& \includegraphics[width=0.4\textwidth]{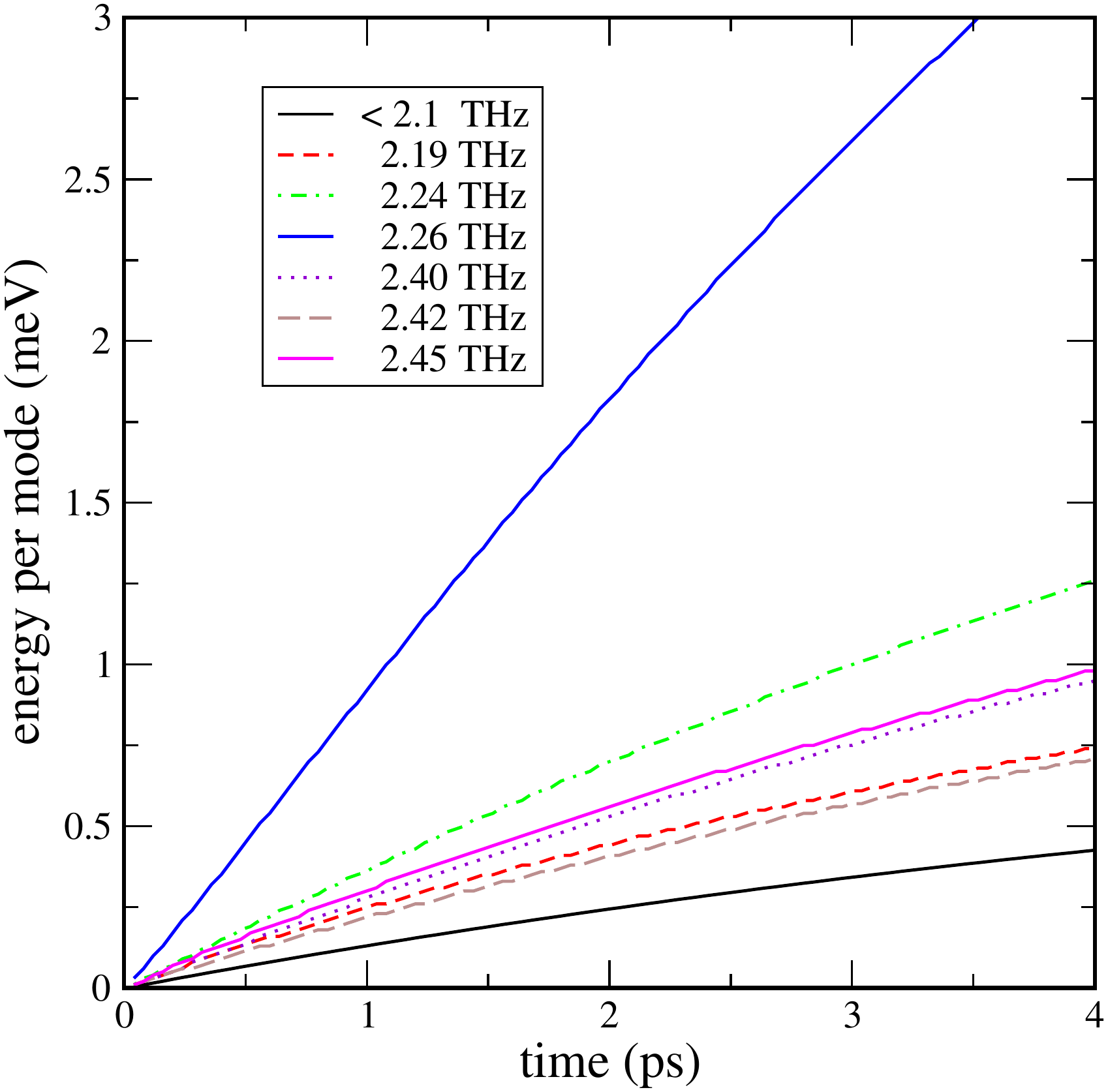} \\ 
(b) & \\ 
& \includegraphics[width=0.4\textwidth]{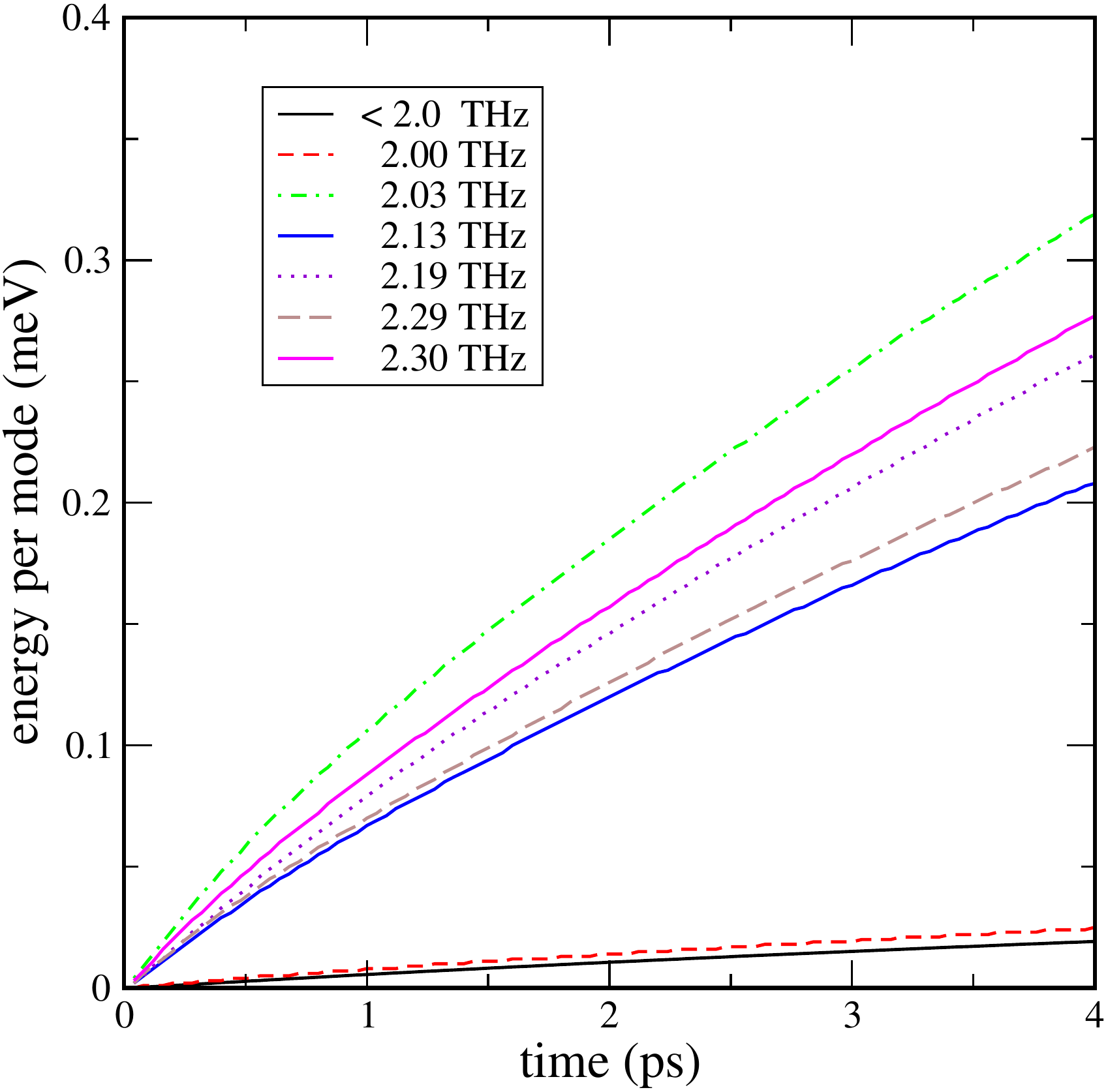}
\end{tabular}
\caption{Temporal evolution of the energy in the highest-lying phonon modes of Pb for (a) 4~ML and (b) 5~ML of Pb/Si(111) films, as obtained from the time-integrated quantities $e_I(t)$ of eq.~(\ref{eq:eI}). 
The strongest phonon excitation is in one particular mode corresponding to surface-normal vibrations of the Pb atoms of the two topmost layers relative to each other at 2.26~THz in the 4~ML film and at 2.03~THz in the 5~ML film, respectively. }
\label{fig:esume}
\end{figure}

Eventually, we comment on the possibility to directly observe the excitation of the lattice 
experimentally. The excitation of low-lying Pb modes results in irregular displacements of the atomic positions on medium to large length scale that can be followed in the Debye-Waller factor of a time-resolved electron diffraction experiment. Preliminary experimental data for Pb films \cite{ZhSt13} indicate that the low-energy phonons are indeed getting excited on the time scale of a few picoseconds. 
Interestingly, recent diffraction experiments were able demonstrate the mode-selective energy transfer to phonons even for bulk materials such as aluminium \cite{Waldecker16} or nickel \cite{Maldonado:2019}. 
Apparently, 
phonon excitation by strongly excited charge carriers opens up the exploration of a new class of non-equilibrium phenomena, not only in bulk metals, but also in semiconductors \cite{SaCh17} and nanostructures. Due to the widely different timescales of e-ph interaction and phonon-phonon interaction, the non-equilibrium distribution of phonons created by the hot carriers can persist over a time span of several picoseconds. Thus, non-thermal phonon distributions could be a more wide-spread phenomenon than previously thought, and further research along these lines could be fruitful.  

\section{Conclusion}
Simulations of electronic relaxation have been performed for a realistic  system, metallic multilayer Pb films on Si(111), with the help of a parameter-free approach based on density functional theory. Electron-electron (e-e) and electron-phonon (e-ph) scattering were both included in the master equation. 
Not surprisingly, e-e scattering was found to dominate over e-ph scattering for short times and highly excited electrons more than 0.5~eV above the Fermi level.  Our simulation results thus justify the neglect of e-ph scattering in the analysis of experimental data \cite{Kirchmann2010} in this regime. The simulated lifetimes  of 101~fs and 21~fs for the quantum well states at 0.58~eV and 1.21~eV, respectively, are in reasonable agreement with the experimental findings. 
The importance of e-ph scattering shows up in the simulations at lower electron energies where the e-e scattering rate decreases strongly.  
Taking the fastest e-ph relaxation found in the present simulations with a characteristic time of 350~fs as a marker for the cross-over between e-e and e-ph scattering, we conclude that e-ph scattering significantly contributes to the relaxation of electrons in Pb at energies below 0.3~eV. Indeed, the simulations show a population pile-up around 0.3~eV due to the cross-over of the e-e and e-ph scattering time scales at this energy. 
After 300~fs, it is fair to describe the electronic population by a thermal distribution, however, only up to an excess population of electrons getting stuck in the 'phonon bottleneck'. Remarkably, this does by no means imply that the phonon populations could be described by a temperature as well. Contrarily, the simulations show that even up to 4~ps after excitation high-frequency surface vibrational modes are preferentially excited by strongly mode-selective phonon emission. 

In summary, our simulations enable us to disentangle the contributions of e-e and e-ph scattering at short times, $<0.3$~ps after optical excitation. 
Although a first glance at the data is compatible with electronic thermalization by e-e scattering, a contribution of e-ph scattering can be observed already in this early stage, in particular at low electron energies. The phonon system requires much longer (several picoseconds) to equilibrate. Additional simulations beyond the scope of the present work are desirable to gain an improved understanding of the energy transfer between the two subsystems in the later stages of relaxation. 

\section*{Acknowledgments}
We thank U. Bovensiepen, M. Horn-von Hoegen and K. Sokolowski-Tinten for useful discussions. 
Financial support within SFB 1242 "Non-equilibrium dynamics of condensed matter in the time domain" funded by Deutsche Forschungsgemeinschaft (DFG), project number 278162697, is gratefully acknowledged. 
We gratefully acknowledge the computing time granted by the Center for Computational Sciences and Simulation (CCSS) of the University of Duisburg-Essen and provided on the supercomputer magnitUDE (DFG Grant No. INST 20876/209-1 FUGG and INST 20876/243-1 FUGG) at the Zentrum f{\"u}r Informations-und Mediendienste (ZIM).

\vfill 

\pagebreak[4]
\begin{widetext}
\appendix
\section{Deformation potential theory of electron-phonon scattering}

The derivation of the form of the electron-phonon matrix element, eq.~(\ref{eq:Dnm}), starts from the Hamiltonian $H_{\rm ep}$ brought to its real-space representation, 
\begin{equation}
\label{eq:HintRealSpace}
H_{\rm ep}(\vec r) =  \sum_{I \vec{Q}}   \Delta V_{I \, \vec Q}(\vec r)
\end{equation}
with 
\begin{equation}
\label{eq:DeltaV} 
\Delta V_{I \, \vec Q}(\vec r) = 
\sqrt{ \frac{\hbar  v_{\mathrm{atom}}^{\mathrm{Pb}} }{2\Omega_{I \vec{Q} }M_{\rm Pb} } }
\left( b^{\dagger}_{I -Q} e^{-i \vec{Q} \vec{ r}} + b_{I Q} e^{i \vec{Q} \vec{r} } \right)
\end{equation}
The physical interpretation of $\Delta V_{I \, \vec Q}(\vec r)$ 
is the additional perturbing potential in the Kohn-Sham equation due to the presence of a phonon of mode $I$ and wave vector $\vec{Q}$. 
The vector notation refers to vectorial quantities in three-dimensional (3D)  space. 

In a very general setting, Fermi's golden rule requires us to calculate matrix elements of the form 
\begin{equation}
\langle \Psi_{ n' \mathbf{k'}}(\vec r) |  \Delta V_{I \, \vec Q}(\vec r) | \Psi_{ n \mathbf{k}}(\vec r)  \rangle 
\end{equation}
involving 3D real-space integration. We now switch to the situation of interest, a two-dimensional film with periodic boundary conditions in the $x,y$ coordinates. 
Within deformation potential theory~\cite{Resta}, the integration over the coordinates $x$ and $y$ in the film plane are carried out to yield a (spatially independent) shift of the Kohn-Sham eigenvalue, denoted by $D_{n\mathbf{k}, \, I \, \mathbf{Q} }$, multiplied by a $\delta$-symbol of parallel momentum conservation, 
\begin{equation}
D_{n\mathbf{k}, \, I \, \mathbf{Q} } \, \delta_{\mathbf{k} - \mathbf{k'}, \pm \mathbf{Q} }
\end{equation}
The bold symbols denote two-dimensional vectors in the film plane; the '$+$' sign refers to absorption, the '$-$' to emission of a phonon of wavevector $\mathbf{Q}$. Technically speaking, the deformation potentials are obtained in the following way: 
A phonon eigenvector labeled $I$, as obtained from the {\sc phonopy} code, is scaled with $\sqrt{M_{\rm Pb}}$ and then added or subtracted from the Cartesian positions of the Pb atoms in the unit cell. For the two geometries obtained in this way (called $+$ and $-$), static DFT calculations are carried out, yielding Kohn-Sham energy eigenvalues $\varepsilon^+_{n \mathbf{k} \, I}$ and $\varepsilon^-_{n \mathbf{k}, I}$. The eigenvalue shift determines the deformation potential via 
\begin{equation} 
\frac{1}{2} \bigl( \varepsilon^+_{n \mathbf{k} \, I } - \varepsilon^-_{n \mathbf{k} \, I} \bigr) =:  \sqrt{ \frac{\hbar}{\Omega_{I 0 }M_{\rm Pb}} } D_{n\mathbf{k}, \, I 0}
\end{equation} 
Note that the (optical) deformation potential $D_{n\mathbf{k}, \, I 0}$ has by definition the physical unit of energy/length.

Now we return to the evaluation of the matrix element in eq.~(\ref{eq:DeltaV}). We work out the equations for general wavevectors $\mathbf{k}$ and  $\mathbf{k}'$ 
and will enforce momentum conservation in a later step. 
After squaring the matrix element to obtain the expression for the rate constant, we can write 
\begin{equation}
\label{eq:intermediate1}
|{\bf D}_{n\mathbf{k},\,I\mathbf{Q}}^{ n'\mathbf{k}' }|^2  = \frac{\hbar v^{\mathrm{Pb}}_{\mathrm{atom}} }{2\Omega_{I 0}M_{\rm Pb} } \left| D_{n\mathbf{k}, \, I}  \langle \phi_{ n' \mathbf{k}' }(z) | e^{i Q_z z} | \phi_{ n \mathbf{k}}(z) \rangle  \right|^2
\end{equation}
Here, the constant eigenvalue shift has been taken out of the brackets, and  $\phi_{ n \mathbf{k}} (z)$ denotes the Bloch-periodic part of the full 3D wavefunction $\Psi_{ n \mathbf{k}}(\vec r)$ suitably averaged over $x$ and $y$. 
This expression requires to evaluate integral(s) over the remaining $z$ coordinate that denotes  the spatial direction normal to the film in which the electrons are confined,  
\begin{equation} 
|{\bf D}_{n\mathbf{k},\,I\mathbf{Q}}^{ n'\mathbf{k}' }|^2  = | D_{n\mathbf{k}, \, I}|^2 \frac{\hbar v^{\mathrm{Pb}}_{\mathrm{atom}} }{2\Omega_{I 0}M_{\rm Pb} }  \int \! dz' \! \int \! dz \, \left( J^{ n'\mathbf{k}' }_{n \,  \mathbf{k} }(z') \right)^* J^{ n'\mathbf{k}' }_{n \, \mathbf{k}} (z) 
\label{eq:twofactors}
\end{equation}
where the definition 
\begin{equation}
J^{ n'\mathbf{k}' }_{n \, \mathbf{k}} (z) = \phi^*_{ n' \mathbf{k}' } (z) e^{i Q_z z} \phi_{ n \mathbf{k}} (z) \, .
\label{eq:defJ}
\end{equation}
has been used. To obtain the overall scattering rate due to one specific phonon mode $I$, the summation over $\vec Q$ in eq.~(\ref{eq:HintRealSpace})  needs to be carried out, or  equivalently an integration over the Brillouin zone of the phonon. Therefore the expression for the rate will include 
\begin{equation}
\sum_{Q_x ,Q_y} \int \frac{dQ_z}{2 \pi}  D_{n\mathbf{k}, I \mathbf{Q} } \, \delta_{\mathbf{k} - \mathbf{k'}, \mathbf{Q} } \int \! dz' \int dz \, \left( J^{ n'\mathbf{k}' }_{n \, \mathbf{k} }(z') \right)^* J^{n'\mathbf{k}' }_{n \, \mathbf{k}} (z)  \, .
\end{equation}
If we now assume that the deformation potential $D$ is only weakly dependent on $\mathbf{Q}$ (which is reasonable for optical phonons in a large supercell), 
we can take it out of the sums and integral, set $\mathbf{Q}=0$ and $D_{n\mathbf{k}, I 0}  = : D_{n\mathbf{k}, I}$,  and perform the integration over $Q_z$ prior to the integration over $z$ and $z'$. 
Motivated by this procedure, we define a quantity (cf. the analogous case for a 1D wire, rather than a 2D film, in the appendix of Ref.~\onlinecite{Ramayya_JAP2008}) 
\begin{equation}
\label{eq:JJ}
 I_{n \, \mathbf{k}}^{ n' \mathbf{k}' } = \int \! dz' \int dz  \int \! \frac{dQ_z}{2\pi}  \, \left( J^{ n'\mathbf{k}' }_{n \, \mathbf{k} }(z') \right)^* J^{n' \mathbf{k}' }_{n \, \mathbf{k}} (z)
\end{equation}
As a consequence of the definition of $J^{ n'\mathbf{k}' }_{n \, \mathbf{k}}$ in eq.~(\ref{eq:defJ}), it follows that the $Q_z$ integration yields $\delta(z-z')$, and the two factors with arguments $z$ and $z'$ in the expression (\ref{eq:twofactors}) turn out to be complex conjugates of each other. 
Performing the integral over $z'$ together with $\delta(z-z')$ leaves us with a single integral over $z$, but now with the squared moduli of the wavefunction in the integrand. 
In the numerical implementation, we not only integrate over $z$, but simultaneously perform the spatial average over the supercell by integrating over $x$ and $y$ and dividing the result by the area of the supercell $A_{\mathrm{supercell}}$:
\begin{equation}
      I_{n\,\mathbf{k}}^{ n' \mathbf{k}' } = A_{\mathrm{supercell}}^{-1} \iiint\!dx\,dy\,dz\ |\Psi_{ n'\mathbf{k}' }(x,y,z)|^2 |\Psi_{ n\mathbf{k} }(x,y,z)|^2.
\end{equation}
For this expression to be valid, the  squared wavefunctions must be normalized such that
$$
\iiint\!dx\,dy\,dz\,  |\Psi_{ n\mathbf{k} }(x,y,z)|^2 = A_{\mathrm{supercell}} \, ,
$$
in other words, $\Psi$ must have the physical unit length$^{-1/2}$. 
Finally, by inserting the expression (\ref{eq:JJ}) for the double integration in eq.~(\ref{eq:twofactors}) and properly normalizing to $ A_{\mathrm{supercell}}$, we arrive at the result given in eq.~(\ref{eq:Dnm}).

In practice, the squared wavefunctions $| \Psi_{ n\mathbf{k} }(x,y,z)|^2$ are obtained from the DFT calculations, using the PARCHG keyword of the VASP code.
To reduce the sheer number of calculations, we calculate $I_{n \, \mathbf{k}}^{ n' \mathbf{k}' }$ explicitly for all combination of band indices $n$ and $n'$ 
of the bands in the energy range $[E_F, E_F+2\mathrm{eV}]$, and for the three combinations of the two $\mathbf{k}$-vectors, $\mathbf{k_x} = (1/4,0,0)$ and $\mathbf{k_y} = (0,1/4,0)$. For other combinations of $\mathbf{k}$-vectors, we interpolate using the angle $\theta$ between 
the wave vectors as variable, defined via the scalar product  $\cos \theta = \frac{  \mathbf{k} \mathbf{k'} }{k k'}$, and obtain  
      \begin{align*}
         I_{n \, \mathbf{k}}^{n'\mathbf{k}' }  = 
         \begin{cases}
          I_{n\, \mathbf{k_x}}^{n'\mathbf{k_x}}  \cos 2 \theta +  I_{n\,\mathbf{k_x}}^{n' \mathbf{k_y} } \sin 2 \theta, &{\rm if} \cos 2 \theta \ge 0 \\[+7pt]
          -  I_{n\,\mathbf{k_y}}^{ n'\mathbf{k_y} }   \cos 2 \theta + I_{n\,\mathbf{k_x}}^{n' \mathbf{k_y} } \sin 2 \theta, &{\rm if} \cos 2 \theta < 0 
        \end{cases}
      \end{align*}

\section{Phonon heat bath}

Already from inspecting the numerical values of the deformation potential $D_{n \mathbf{k}, \, I}$, it becomes clear that the coupling of an excited electron to a phonon can be  highly phonon-mode specific. In particular, coupling to the highest-lying Pb modes is strong.        
We therefore use an independent heat bath for each the $N_{\rm opt} = 6$ highest-lying Pb modes. 
For the 4ML Pb/Si(111) film, these six phonons (surface and interface modes) of the Pb layer are in the 2.1 to 2.5~THz range (modes labelled 44, 45, 46, 47, 48 and 49 in Ref.~\onlinecite{ZaKr17}). 
For the 5ML Pb/Si(111) film, the analogous modes are found at somewhat lower frequencies, 
in the 2.0 to 2.3~THz range. 
All the other, lower-lying modes are taken to constitute a common acoustic phonon bath with temperature $T_{0}$. 
In addition, a constant lattice temperature $T_{\mathrm{sub}}$ of the Si substrate acting as a heat sink  is part of the description of the vibrational system. 

As initial condition, all temperatures are set to a base temperature of 100~K at the beginning of the simulation. 
In general, the population $n_I(t)$ of a phonon mode $I$ varies slowly on the scale of electronic relaxation. 
Yet we allow for energy exchange between the various heat baths.  
Previous simulations \cite{Sakong2013} using classical molecular dynamics showed that the surface phonon in a Pb/Si(111) monolayer is damped due to mode conversion on a time scale of $\tau_{\rm conv}$ = 30~ps. In the present simulation, we use this time constant to couple each of the baths at temperature $T_I(t)$ to the common acoustic bath at $T_0(t)$. 
Moreover, the acoustic bath may transfer energy to the Si substrate lattice on an even longer time scale  
$\tau_{\rm sub} = 160$~ps. This value was adopted from measurements observing the equilibration of Pb films with the Si substrate by Witte {\em et al.} \cite{WiFr17}.  
The above considerations lead to the following equations describing the evolution of lattice temperatures:  
\begin{eqnarray}
          \frac{dT_I(t)}{dt} &=& \frac{ T_0(t) - T_I(t) }{\tau_{\rm conv} } +  \frac{e_I(t)}{c_{V}^{\rm opt} },  \label{eq:optBath}\\
          \frac{dT_0(t)}{dt} &=& \sum_{I= N_{\rm acu} + 1}^{N_{\rm acu}+ N_{\rm opt}}  \frac{(T_I(t) - T_0(t) )c_{V}^ {\rm opt} }{\tau_{\rm conv} \, c_{V}^{\rm acu} } - \frac{T_0(t) - T_{\rm sub} }{\tau_{\rm sub}} +  \frac{e_0(t)}{c_{V}^{\rm acu} } .
          \label{eq:acuBath} 
\end{eqnarray}
The quantities $c_{V}^{\rm acu}$ and $c_{V}^{\rm opt}$ denote the partial heat capacities of the acoustic phonon bath, and of one optical mode, respectively. 
In the 'classical' approximation, i.e. when the equipartition theorem holds, we have $c_{V}^ {\rm opt} =  k_B$ and $c_{V}^{\rm acu} = N_{\rm acu} k_B$. In the present cases, $N_{\rm acu} = 42$ for the 4~ML slab and $N_{\rm acu} = 54$ for the 5~ML slab.
The rates of energy transfer $e_I(t)$ from the electronic system into the heat bath are calculated simultaneously while solving the master equation (\ref{eq:Master}) and eq.~(\ref{eq:rateEPH}). They define the inhomogeneous terms on the left hand side of eq.s~(\ref{eq:optBath}) and (\ref{eq:acuBath}), 
\begin{equation}
\label{eq:eI}
e_I(t) = 2\pi \sum_{\substack{n,\,m,\\ \mathbf{Q},\,\pm}}  \pm \hbar\Omega_{I\mathbf{Q}}  \, \left|{{\bf D}_{n\mathbf{k},\,I\mathbf{Q}}^{m,\,\mathbf{k}-\mathbf{Q}}}\right|^2 \left(n_{I\mathbf{Q}} + \frac{1}{2} \pm \frac{1}{2}\right)\,\delta\big(\varepsilon_{n\mathbf{k}} - \varepsilon_{m\mathbf{k}-\mathbf{Q}} \pm \hbar\Omega_{I\mathbf{Q}} \big)\,  \big( f_{m,\,\mathbf{k}-\mathbf{Q}}  -  f_{n, \, \mathbf{k}} \big) 
\end{equation}
with $I=N_{\rm acu}+1, \ldots N_{\rm acu}  + N_{\rm opt}$, and 
$
e_0(t) =   \sum_{I=1}^{N_{\rm acu}} e_I(t) \, .
$
The time integral of these quantities yields the energy transfer to each vibrational mode displayed in Fig.~\ref{fig:esume}.  
\medskip

\begin{figure}[tbp]
\begin{center}
\includegraphics[width=0.35\textwidth]{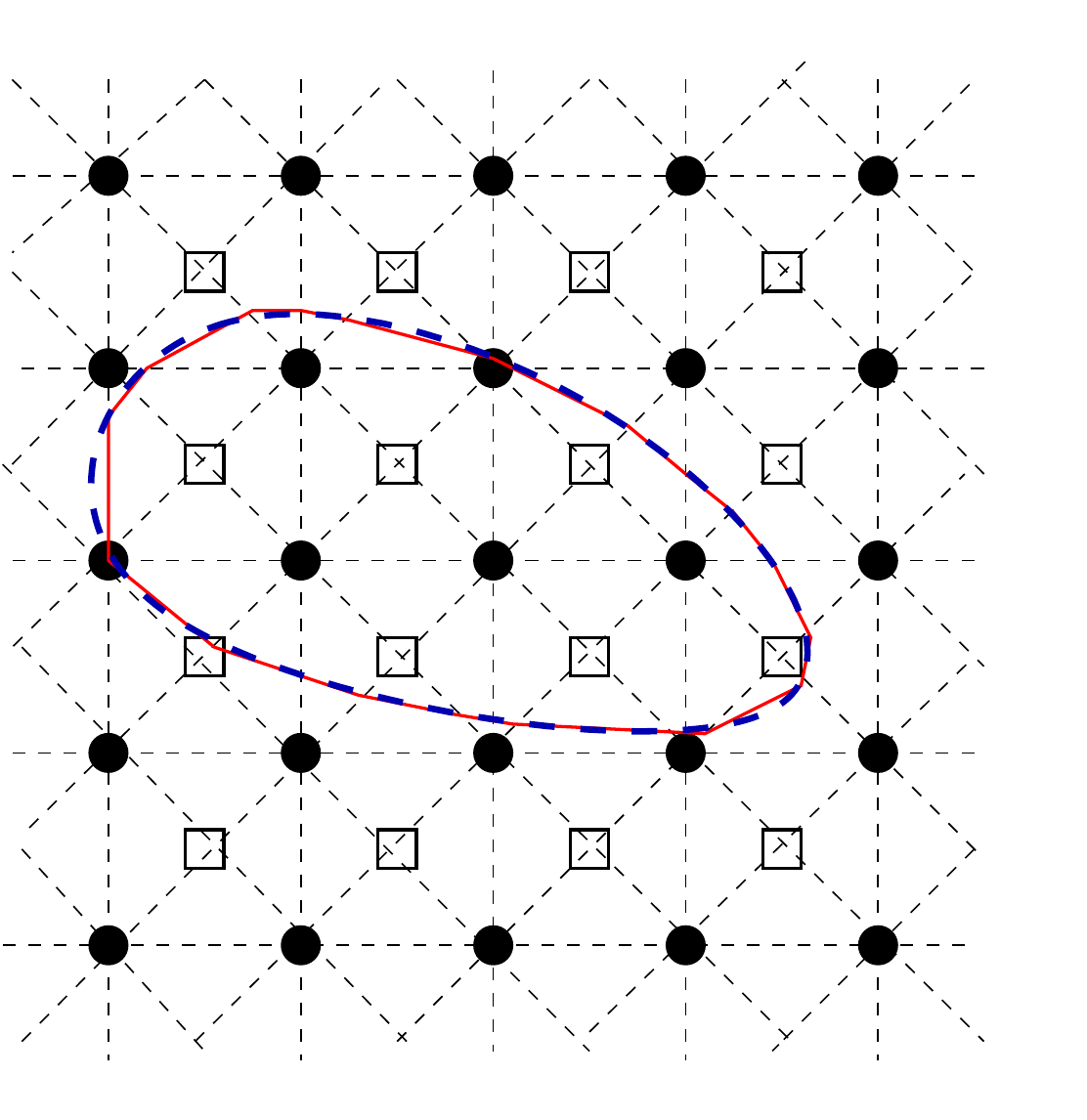}
\begin{center}
\caption{The figure illustrates schematically how energy conservation is imposed when evaluating Brillouin zone integrals. The support points of the population data vector $f_{n\mathbf{k}}$ (full circles) and the additional mid-point grid (empty squares) define a tessellation of the Brillouin zone by triangles (thin dashed lines). 
The thick dashed blue line represents the solution of eq.~(\ref{eq:eneConserved}) expressing energy conservation, and the full red line segments are the approximative solutions constructed in each triangle by local linearization of the bands $\varepsilon_{n\mathbf{k}}$.
} 
\label{fig:grid}
\end{center}
\end{center}
\end{figure}     

\section{Numerical methods}

The electronic Brillouin zone is sampled on a $32 \times 32 \, \mathbf{k}$-space grid. Due to symmetry, only one half of the Brillouin zone with grid points $k_y \ge 0$ needs to be stored, resulting in a $32 \times 17$ grid of data points. For storing the electronic population data vector $f_{n\mathbf{k}}$, the grid is refined by a second grid defined by the mid-points of the first grid. The Master equation is integrated with a time step of $\Delta t = 0.04$~fs  by iterating between the two grids. 
For evaluating the energy-conserving $\delta$-function in eq.~(\ref{eq:rateEPH}), it is transformed into a $\delta$-function in $\mathbf{k}$-space using the relation 
\begin{equation}
\delta\big(\varepsilon_{n\mathbf{k}} - \varepsilon_{m\mathbf{k}-\mathbf{Q}} \pm \hbar\Omega_{I0} \big) = 
\left| \frac{d( \varepsilon_{n\mathbf{k}} - \varepsilon_{m\mathbf{k-}\mathbf{Q}} )}{d\mathbf{k}} \right|^{-1} 
\delta^{(2)}(\mathbf{K}(n,m,I) - \mathbf{k} + \mathbf{Q} ) 
\label{eq:delta2} 
\end{equation}
where $\mathbf{K}(n,m,I)$ denotes the one-dimensional solutions of the equation
\begin{equation}
\label{eq:eneConserved}
\varepsilon_{n\mathbf{k}} - \varepsilon_{m\mathbf{K}(n,m,I)} = \pm \hbar \Omega_{I 0}
\end{equation}
for fixed $n, \, m, \, I$ and $\mathbf{k}$. The solutions are constructed from straight line segments  in each triangle in a tessellation of the Brillouin zone defined by the two interpenetrating grids. In each triangle, the $\varepsilon_{n\mathbf{k}}$ are approximated as piecewise bi-linear functions of $k_x$ and $k_y$.  
Eventually, the integration of the $\delta^{(2)}$-function in eq.~(\ref{eq:delta2}) is numerically represented by the arc length (in $\mathbf{k}$-space) of the solution, approximated by the cumulated lengths of the line segments, as illustrated in Fig.~\ref{fig:grid}.   
The right hand side of the Master equation is thus made up by the sum over $m,\, I$, all grid points and pertaining line segments, in total $\sim 10^6$ terms. 

\end{widetext}

\providecommand{\noopsort}[1]{}\providecommand{\singleletter}[1]{#1}%

\end{document}